\documentclass[a4paper, reprint, superscriptaddress, amsmath, amssymb, aps, prl, twoside]{revtex4-2}
\usepackage[english]{babel}
\usepackage[utf8]{inputenc}
\usepackage[colorinlistoftodos, color=green!40, prependcaption]{todonotes}
\usepackage{amsthm}
\usepackage{mathtools}
\usepackage{physics}
\usepackage{xcolor}
\usepackage{graphicx}
\usepackage[left=23mm,right=13mm,top=35mm,columnsep=15pt]{geometry} 
\usepackage{adjustbox}
\usepackage{placeins}
\usepackage[T1]{fontenc}
\usepackage{lipsum}
\usepackage{csquotes}
\usepackage[pdftex, pdftitle={Article}, pdfauthor={Author}]{hyperref} % For hyperlinks in the PDF
\usepackage{upgreek}
\begin{document}
\title{Scanning cavity microscopy of a single-crystal diamond membrane}

\author{Jonathan Körber}
\affiliation{ 
Physikalisches Institut, Karlsruhe Institute of Technology (KIT), Wolfgang-Gaede Str. 1, 76131 Karlsruhe, Germany}

\author{Maximilian Pallmann}
\affiliation{ 
Physikalisches Institut, Karlsruhe Institute of Technology (KIT), Wolfgang-Gaede Str. 1, 76131 Karlsruhe, Germany}

\author{Julia Heupel}
\affiliation{Institute of Nanostructure Technologies and Analytics (INA), Center for Interdisciplinary Nanostructure Science and Technology (CINSaT), University of Kassel, Heinrich-Plett-Straße 40, 34132 Kassel, Germany}

\author{Rainer Stöhr}
\affiliation{3rd Institute of Physics, University of Stuttgart, Pfaffenwaldring 57, 70569 Stuttgart, Germany}

\author{Evgenij Vasilenko}
\affiliation{ 
Physikalisches Institut, Karlsruhe Institute of Technology (KIT), Wolfgang-Gaede Str. 1, 76131 Karlsruhe, Germany}
\affiliation{Institute for Quantum Materials and Technologies (IQMT), Karlsruhe Institute of Technology (KIT), Herrmann-von-Helmholtz Platz 1, 76344 Eggenstein-Leopoldshafen, Germany}

\author{Thomas Hümmer}
\affiliation{Faculty of Physics, Ludwig-Maximilians-University (LMU), Schellingstr. 4, 80799 Munich, Germany}

\author{Larissa Kohler}
\affiliation{ 
Physikalisches Institut, Karlsruhe Institute of Technology (KIT), Wolfgang-Gaede Str. 1, 76131 Karlsruhe, Germany}

\author{Cyril Popov}
\affiliation{Institute of Nanostructure Technologies and Analytics (INA), Center for Interdisciplinary Nanostructure Science and Technology (CINSaT), University of Kassel, Heinrich-Plett-Straße 40, 34132 Kassel, Germany}

\author{David Hunger}
\affiliation{ 
Physikalisches Institut, Karlsruhe Institute of Technology (KIT), Wolfgang-Gaede Str. 1, 76131 Karlsruhe, Germany}
\affiliation{Institute for Quantum Materials and Technologies (IQMT), Karlsruhe Institute of Technology (KIT), Herrmann-von-Helmholtz Platz 1, 76344 Eggenstein-Leopoldshafen, Germany}
\email{david.hunger@kit.edu}

\date{\today} 

\begin{abstract} 
Efficient optical interfacing of spin-bearing quantum emitters is a crucial ingredient for quantum networks. 
A promising route therefore is to incorporate host materials as minimally processed membranes into open-access microcavities: 
it enables significant emission enhancement and efficient photon collection, minimizes deteriorating influence on the quantum emitter, and allows for full spatial and spectral tunability.
Here, we study the properties of a high-finesse fiber Fabry-Pérot microcavity with integrated single-crystal diamond membranes by scanning cavity microscopy. We observe spatially resolved the effects of the diamond-air interface on the cavity mode structure: a strong correlation of the cavity finesse and mode structure with the diamond thickness and surface topography, prevalent transverse-mode mixing under diamond-like conditions, and mode-character-dependent polarization-mode splitting. Our results reveal the influence of the diamond surface on the achievable Purcell enhancement, which helps to clarify the route towards optimized spin-photon interfaces. 
\end{abstract}


\maketitle

\section{Introduction} \label{sec:introdcution}
Quantum networks built from individual optically addressable spins in solids interfaced with single photons \cite{Awschalom2018,Atatuere2018} promise a variety of emerging applications, ranging from secure communication over large distances to distributed quantum computing, which could become the basis of a future quantum internet \cite{Wehner2018}. A key building block for this is an efficient interface between the spin and photons to enable deterministic transfer of quantum states between stationary and flying qubits \cite{Johnson2017,Borregaard2019,Janitz2020,Ruf2021b}. The most powerful approach in this respect is to couple quantum emitters to optical microcavities and harness the Purcell effect to enhance the emission \cite{Purcell,Reiserer2015}. This reshapes the emission pattern into a single, well collectable mode, increases the emission fraction into the coherent zero-phonon line (ZPL), and broadens the transition due to the shortening of the excited state lifetime such that spectral fluctuations can be masked to improve photon indistinguishability.

\begin{figure}
    \includegraphics[width=0.48\textwidth]{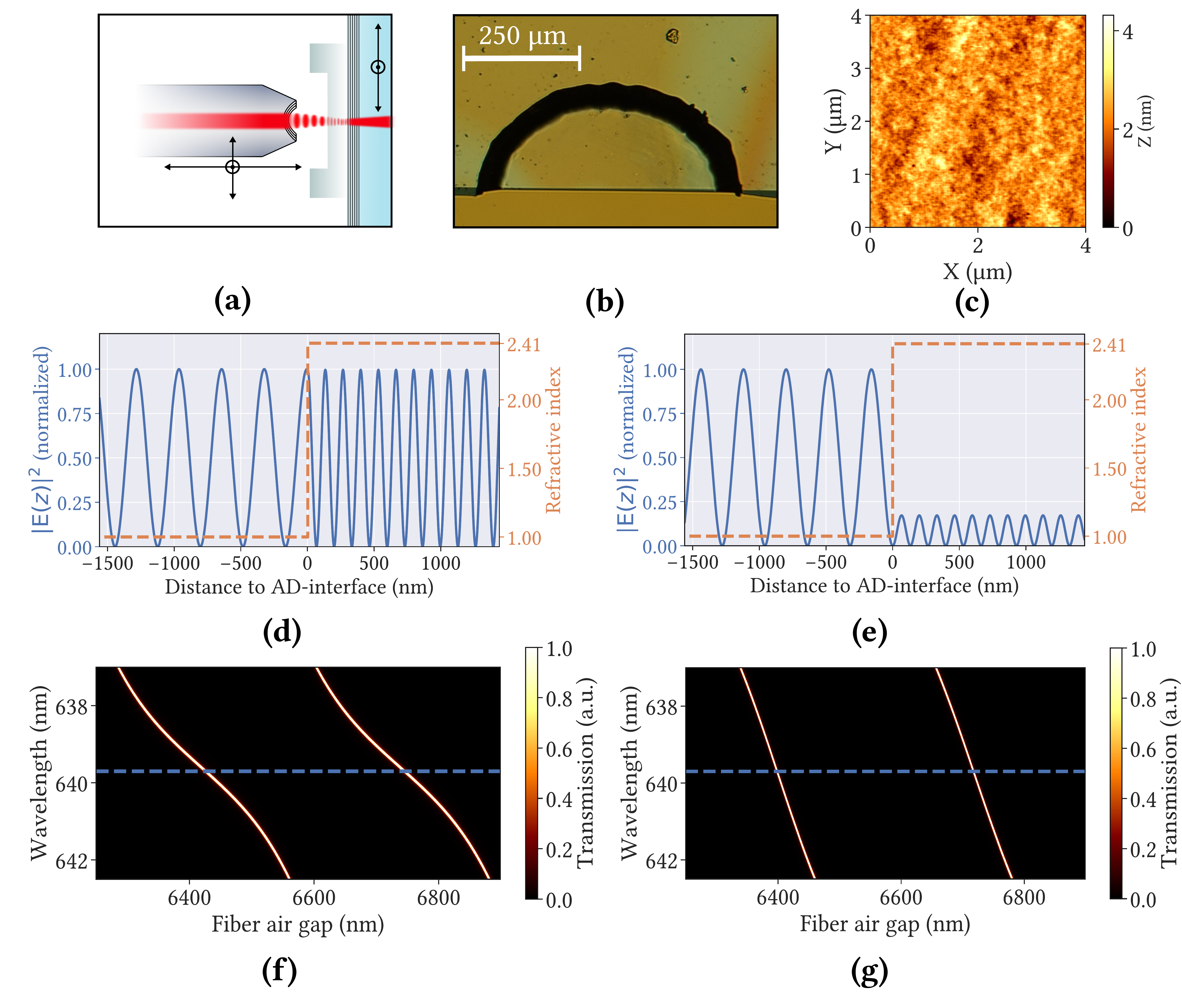}
\caption[]{\textbf{(a)} Schematic drawing of the cavity setup: A planar mirror (blue) carrying a diamond membrane can be moved laterally to select a region of interest. A cavity is formed together with a fiber mirror that can be nano-positioned along three axes to record raster-scanning images. \textbf{(b)} Microscope image of a bonded membrane. A thinned-down part for subsequent characterization lies within the region framed by the semicircle shadow.  \textbf{(c)} AFM measurement of the membrane showing an rms-roughness of $\sigma_{\mathrm{rms}} = 0.4 \, \mathrm{nm}$. 
Simulation of the electric field inside the cavity for a diamond-like \textbf{(d)} and an air-like \textbf{(e)} mode. The refractive index profile (orange dashed line) highlights the air-diamond (AD) interface. \textbf{(f)} and \textbf{(g)}: Simulated cavity resonance frequency as a function of the mirror separation for a diamond thickness of $6035 \, \mathrm{nm}$ (diamond-like) and $6103 \, \mathrm{nm}$ (air-like), respectively. The probe wavelength of subsequent measurements $\lambda = 639.7 \, \mathrm{nm}$  is indicated by a dashed blue line.}
\label{Fig1}
\end{figure}

Nitrogen-vacancy (NV) centers in diamond are a prime candidate material system in this respect that stands out due to its exceptional spin coherence properties, the availability of a nuclear spin quantum register \cite{Bradley2019}, and lifetime-limited optical transitions that permit the generation of spin-photon entanglement \cite{Bernien2013}. Several cavity architectures have been investigated to demonstrate the basic principle of Purcell enhancement of NV center emission \cite{Faraon2011,Faraon2012,Hausmann2013,Albrecht2013,Li2015,Kaupp16,Dolan18}. Due to the high sensitivity of NV centers to fluctuating electric fields, a promising approach to achieve narrow optical transitions inside a cavity is based on open-access Fabry-Pérot micro-cavities with incorporated minimally processed single-crystal membranes 
\cite{Janitz.2015,Bogdanovic2017,Riedel2017,Ruf2019,Heupel.2020,Ruf2021}, see figure~\ref{Fig1}a).
This approach has led to first successful experiments, both with NV \cite{Riedel2017,Ruf2021}, SiV \cite{Haeussler2019,Salz2020} and GeV centers \cite{Jensen2020} in diamond, as well as with rare earth ions in oxide crystals \cite{Merkel2020}. However, to the best of our knowledge, the desired net improvement of the collection of lifetime-limited NV center ZPL photons has not been achieved to date. In earlier work, either the collection efficiency was rather low \cite{Ruf2021}, or the NV emission linewidth was more than two orders of magnitude above the lifetime limit \cite{Riedel2017}.
This emphasizes the further need to understand limitations and improve the experimental realization of such systems.

In this work, we perform a systematic study of the cavity performance in the presence of diamond membranes at room temperature. We use scanning cavity microscopy \cite{Mader2015,Kelkar2015,Huemmer2016} at a high finesse of up to 20000 to probe extended areas of a membrane. We observe a spatial variation of the cavity mode character and identify spatially localized mode mixing \cite{Benedikter.2015,Benedikter.2019}, mostly present in diamond-like regions. We model the observed losses based on surface scattering, absorption, and mirror transmission, and find that we need to assume a larger surface roughness than measured. Additionally, we include bulk absorption in our model and find good agreement with the expected absorption for the used diamond. Furthermore, we observe an increase of the cavity loss for air-like modes with increasing diamond thickness. We suggest that the mismatch between the mode's curved phase front and the planar diamond-air interface can be the origin. Finally, we study the polarization modes of the cavity and observe an increased polarization mode splitting with a dependence on the mode character. This reveals birefringence of the membrane, which we interpret as a signature of the local strain distribution in the material.   

Our results show that mode mixing, wavefront curvature, and polarization mode splitting are important contributions to the performance of cavities with integrated membranes. We analyze how these effects depend on geometrical properties of the membrane and the cavity, providing guidance to achieve optimized spin-photon interfaces.

\section{Methods and Materials} \label{sec:methods}
Our experiments are performed using a fiber-based Fabry-Pérot cavity which is schematically illustrated in figure \ref{Fig1} (a). It consists of a macroscopic plane mirror and a concave mirror that is processed at the end facet of an optical fiber by CO$_2$-laser machining \cite{co2fab2011}. We use two different fiber tips F$_{\mathrm{A}}$ (F$_{\mathrm{B}}$) that show concave profiles with small ellipticity, characterized by radii of curvatures (ROCs) of $33.1 \, (55.2) \, \upmu \mathrm{m}$ along one half axis and $30.5 \, (48.7) \, \upmu \mathrm{m}$ along the orthogonal one. The fiber end facets are coated by ion beam sputtering with a distributed Bragg reflector (DBR) (Laseroptik GmbH, Garbsen, Germany), designed for a center wavelength at $637 \, \mathrm{nm}$. Numerical simulations using a transfer matrix model along with the measured layer thicknesses of the DBR stacks yield transmissions of $52 \, (57) \, \mathrm{ppm}$ for the fiber mirrors at a wavelength of $\lambda = 639.7 \, \mathrm{nm}$ that is used for the experiment. The plane mirror (M$_{\mathrm{A}}$) consists of a superpolished fused silica substrate ($\sigma_{\mathrm{rms}} < 0.2 \, \mathrm{nm}$), coated for a transmission of $57 \, \mathrm{ppm}$ (see the Supplementary Material \cite{supp} for details on the cavity mirrors as well as the expected and measured finesse of the assembled cavities without the presence of a diamond sample. 

We study two CVD-grown single-crystal diamond samples of general grade (\textit{Cornes Technologies, San Jose, CA, USA}) and electronic grade (\textit{element 6}) quality. While most approaches propose the use of ultra pure, electronic grade diamond samples, also samples with a higher natural abundance of nitrogen have yielded NV centers with very promising optical coherence properties recently \cite{orphal-kobin_optically_2022}. Also, we choose here to study one of the two samples being non-optimal in order to clearly identify and quantify limiting factors. Such factors can still affect optimized samples, but would be difficult to differentiate without this comparison. The main sample of this study (general grade quality) is structured using an inductively-coupled plasma reactive ion etching (ICP-RIE) procedure, resulting in a membrane with a minimal thickness of approximately $5 \, \upmu \mathrm{m}$, as described in \cite{Heupel.2020}. Characterization on $4 \times4 \, \mathrm{\upmu m}^2$ sub-regions of the membrane with an atomic force microscope (AFM) - as an example shown in figure \ref{Fig1} (c) - reveals a surface roughness of $\sigma_{\mathrm{rms}} = 0.4 - 0.5 \, \mathrm{nm}$ after the final etching steps. We bond the membranes onto plane mirrors via van der Waals forces. Initial interference fringes observed under a light microscope, indicating an air gap between the membrane and the mirror, broaden during the bonding procedure, and the final result for the general grade sample depicted in figure \ref{Fig1} (b) shows almost no fringes (see \cite{Heupel.2020} for details on the bonding procedure).

We use a custom-developed nanopositioning stage to control the cavity. It can be built cryo-compatible and achieves very high passive stability and fast scanning speed \cite{Casabone2021,Pallmann2022}. In order to select a position on the a sample, the plane mirror can be moved laterally, while the fiber can be scanned additionally along all three dimensions using piezo-electric actuators to perform the scanning cavity measurements. To probe the cavity, we couple light of a tuneable diode laser (TDL) (\textit{Toptica DL pro 637}) at a wavelength of $\lambda = 639.7 \, \mathrm{nm}$ into the cavity fiber and measure the transmission behind the plane mirror using an avalanche photodiode (APD) (\textit{Thorlabs APD130A2/M}) connected to a 12-bit digital storage oscilloscope. To obtain spatially resolved maps, we raster-scan the fiber laterally on an area of about $60\times 60\, \upmu \mathrm{m}^2$ over the sample and modulate the cavity length at each position over multiple free spectral ranges (FSR) at a rate of $200 \, \mathrm{Hz}$. The maximum transmission is recorded at each position, leading to a 2D-image that shows relative changes of the cavity transmission as a function of the position on the membrane, which we refer to as a cavity transmission scan. To obtain the cavity finesse, we modulate the cavity length at a reduced rate of typically $20 \,\mathrm{Hz}$ and probe two consecutive fundamental resonances, which we fit with Lorentzian lines. From this we calculate the finesse from the ratio of the resonance distance and the average full width at half maximum (FWHM). Due to piezo nonlinearity and hysteresis, such measurements have an uncertainty of $\approx 10\%$ unless calibrated with care. For these measurements, we align the polarization of the input light with one of the two cavity polarization modes to probe an isolated resonance.

The presence of the diamond-air interface leads to a hybridized mode structure \cite{Janitz.2015,SuzanneBvanDam.2018, Janitz2020,Heupel.2020}. Depending on the diamond thickness, the cavity standing wave light field can either fulfill the boundary condition for the air gap part where a field node should form at the interface, called an air-like mode, or match for the mode in the diamond part where a field maximum at the interface is present, a so-called diamond-like mode, see Fig.~\ref{Fig1} (d) and (e). Since both conditions cannot be met simultaneously, hybridized modes form, whose dispersion, finesse and loss strongly depend on the mode character \cite{Janitz.2015,SuzanneBvanDam.2018}. For such a hybridized diamond-air cavity, the mode dispersion, i.e. the frequency dependent shift of the cavity resonances, deviates from the linear behavior known from a bare Fabry-Pérot cavity and shows a varying slope for different diamond thickness at a fixed wavelength. As shown in a simulation of cavity resonances in figure \ref{Fig1} (f), the dispersion exhibits the smallest slope for a configuration with predominant diamond-like character, and the steepest slope for a predominant air-like configuration as in figure \ref{Fig1} (g). To maximize the coupling to color centers in the membrane, diamond-like modes are beneficial because they show a larger electric field inside the diamond \cite{SuzanneBvanDam.2018} as it can be seen by comparing figures \ref{Fig1} (d) and (e). However, for this configuration, the scattering loss at the diamond surface becomes maximal, requiring a trade-off depending on the surface roughness level. Furthermore, a recent study observed additional loss associated with diamond-like modes which could not be conclusively explained by surface scattering \cite{Flagan2021}. In general, a comprehensive investigation of the cavity performance and its relation to the diamond surface topography would be desirable, while experiments to date report only punctual measurements of cavity performance on diamond membranes, mostly under air-like conditions.

\section{Results and Discussion} 
\subsection{Mode-character-dependent cavity loss}\label{sec:modecharacter}
We study the spatially varying effect of the diamond membrane on cavity modes by performing laterally resolved cavity transmission measurements as described above. To increase the imaged area, 16 of such scans are performed at neighboring positions on the membrane with slight overlaps at the edges and consecutively stitched together.

\begin{figure}
    \includegraphics[width=0.48\textwidth]{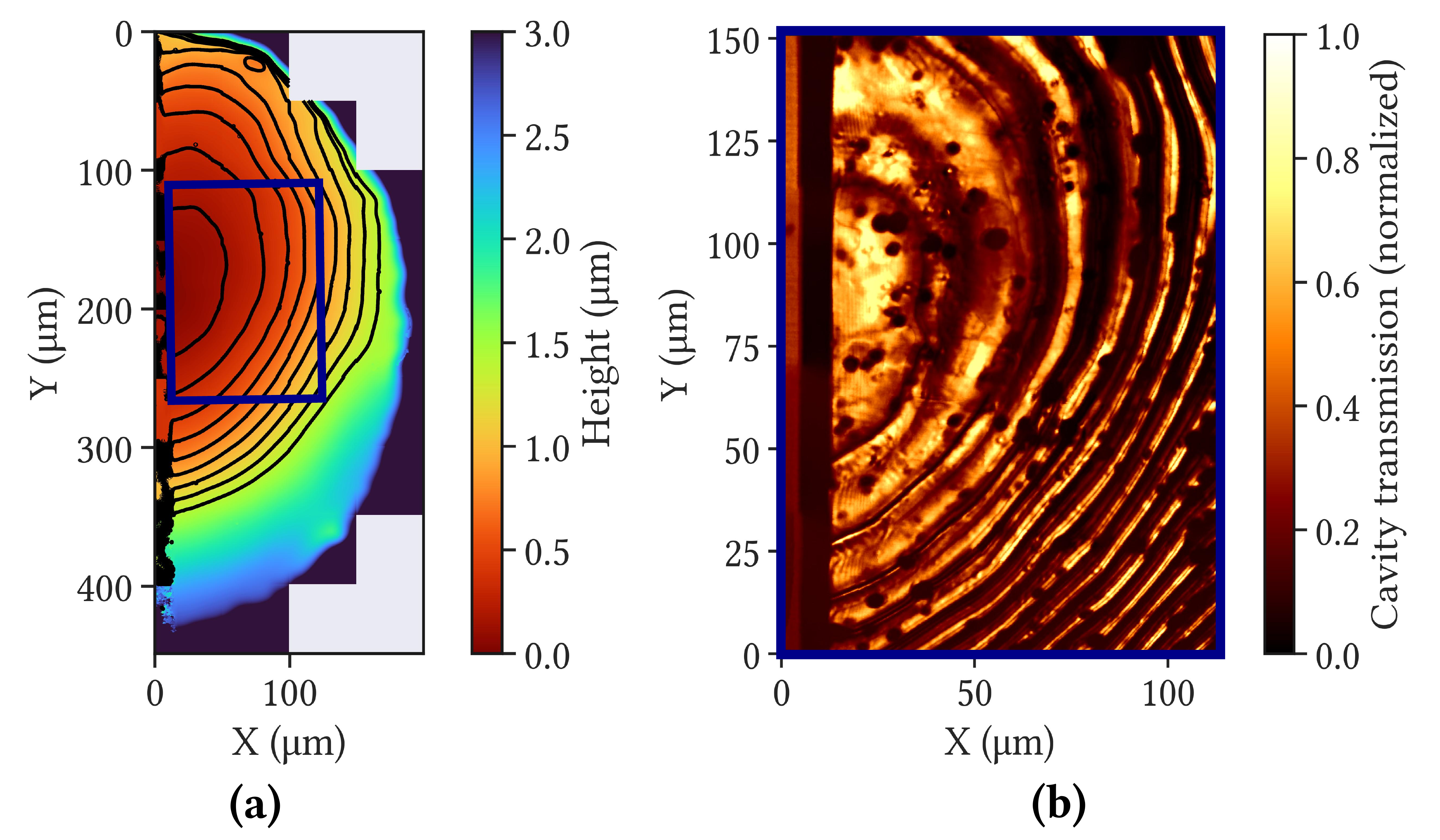}
\caption[]{\textbf{(a)} Surface topography of the etched diamond membrane. The blue frame indicates the region where cavity transmission scans are performed. Black contour lines with a stepsize of 156$\,\mathrm{nm}$ are used to highlight the height profile of the membrane. \textbf{(b)} Cavity transmission map of the blue framed region in (a).}
\label{Fig2}
\end{figure}

The final transmission scan obtained with fiber mirror F$_{\mathrm{A}}$ is shown in figure \ref{Fig2} along with a height map of the membrane taken with a home-built white-light interferometric microscope (WLI). Notably, the cavity transmission in figure \ref{Fig2} (b) exhibits distinct bright and dark fringes that match the shape of the height map in figure \ref{Fig2} (a), evidencing the correlation between the cavity transmission and the membrane thickness. Since the composition of the cavity mode changes with the diamond thickness at a periodicity of $\lambda / \left(2n_{\mathrm{d}}\right)$ \cite{SuzanneBvanDam.2018}, with $n_{\mathrm{d}} = 2.41$ the refractive index of diamond, our measurements reveal the relative change of the cavity transmission as a function of the hybridized mode composition. To compare the thickness change of the diamond membrane obtained from the WLI measurement to the structure observed in cavity transmission, we estimate the region covered by the cavity scans (see Supplementary Material \cite{supp}). This region is indicated by the blue frame in figure \ref{Fig2} (a). For a horizontal line at the central y-position of this region, the WLI image shows a thickness change of $\Delta_{\mathrm{h}}^{\mathrm{WLI}} \approx 780 \, \mathrm{nm}$. Roughly seven bright fringes along the x-axis in the central y position of the cavity scan yield a thickness change of $\Delta_{\mathrm{h}}^{\mathrm{cav}} = (7-1) \cdot \lambda / \left(2n_{\mathrm{d}}\right) \approx 800 \, \mathrm{nm}$, thus showing good agreement with the WLI measurement. 

\begin{figure*}
    \includegraphics[width=0.8\textwidth]{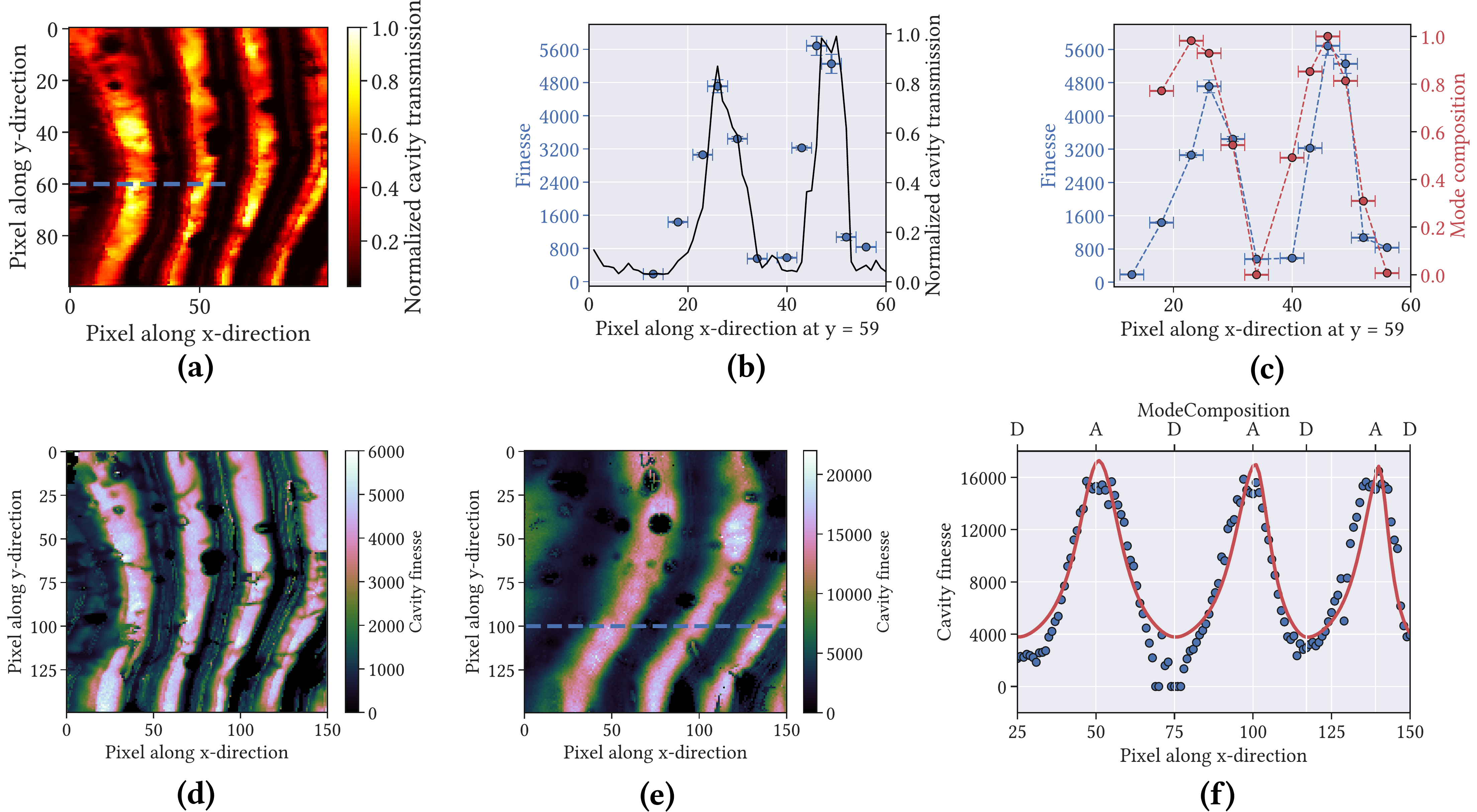}
\caption[]{\textbf{(a)} Cavity transmission scan on the membrane. Each pixel shows the maximum cavity transmission from several fundamental mode orders. \textbf{(b)} Finesse of the cavity for points along the blue line in (a). Each data point shows the average of 50 measurements. The black curve shows the transmission data from (a). \textbf{(c)} Cavity finesse (data from (b)) and composition of the hybridized mode obtained by measuring the mode dispersion for each point. \textbf{(d)} Cavity finesse scan on the same region. \textbf{(e)} Cavity finesse scan using a different cavity fiber mirror achieving a higher finesse on a slightly different position on the membrane. \textbf{(f)} Measured cavity finesse along the dashed blue line in (e). The red line shows a fit.}
\label{Fig3}
\end{figure*}

To understand how the composition of the hybridized cavity mode affects the cavity losses in more detail, we proceed with measuring the cavity finesse and the mode composition for points along a horizontal line, covering several bright and dark fringes in figure \ref{Fig3} (a). As shown in figure \ref{Fig3} (b), the cavity finesse spans values between 200 and 5500 and follows closely the cavity transmission measurements. In order to link each measurement position with the character of the hybridized mode, we investigate the dispersion of the cavity mode. To this end, we couple light from a white-light laser (\textit{Fianium  Whitelase SC450}) into the cavity and measure transmission spectra with a spectrometer (\textit{Andor Shamrock}) for different cavity lengths. The mode character is derived from the slope of the mode dispersion at the wavelength $\lambda = 639.7 \, \mathrm{nm}$ of the probe laser used for the finesse measurements. Here, a steep slope corresponds to a mode with high air-like character and a flat slope to a mode with high diamond-like character \cite{Janitz.2015}. We define the mode character by assigning a value of 1 for the steepest slope (air-like, A) and 0 for the lowest slope (diamond-like, D) with corresponding values in between (see Supplementary Material \cite{supp}).

The results are shown together with a corresponding measurement of the finesse in figure \ref{Fig3} (c). The mode character follows closely the finesse and the transmission of the cavity, proving that the losses of the cavity are dominated by effects associated with the character of the hybridized mode. To evidence the strong correlation between the cavity losses and the mode composition again over a larger region of the membrane, we scan the fiber similar to the lateral transmission measurements across the membrane and measure the finesse of the cavity at each position. The resulting finesse scan in figure \ref{Fig3} (d) confirms the correlation between cavity finesse and mode composition obtained along the single line described above.

We repeat the measurement with a second fiber mirror F$_{\mathrm{B}}$ since we observed high losses for the first fiber mirror already without diamond membrane. Taken at a slightly different position, the measurement shown in figure \ref{Fig3} (e) again reflects the shape of the membrane. The finesse spans values between $\sim 3000$ and slightly above $20,000$, which remains lower than the value observed for the cavity without membrane $\mathcal{F}^{\mathrm{bare}}_{\mathrm{max}} = 32,400$ (see Supplementary Material \cite{supp}). 
To understand the contribution of different sources of loss, we analyze a line cut through the data shown in figure \ref{Fig3} (e) and fit it with a model containing mode-dependent mirror transmission, scattering losses at the diamond-air interface, and absorption from the diamond \cite{SuzanneBvanDam.2018}, see figure \ref{Fig3} (f). Therefore, we describe the finesse $\mathcal{F} = 2 \pi / \mathcal{L}_{\mathrm{eff}}$ with the effective losses \cite{SuzanneBvanDam.2018}
\begin{equation}
\mathcal{L}_{\mathrm{eff}} = \frac{E_{\mathrm{max,a}}^2}{n_{\mathrm{d}}E_{\mathrm{max,d}}^2} \mathcal{L}^{f} + \mathcal{L}^{m} + \mathcal{L}_{\mathrm{scat}} (\sigma_{\mathrm{rms}}) + \mathcal{L}_{\mathrm{abs}} (\alpha_{\mathrm{d}})
\end{equation}
Here, $\mathcal{L}^{m}$ ($\mathcal{L}^{f}$) are the losses due to transmission at the planar (fiber) mirror that we extract by a transfer matrix model using the measured DBR-layer thicknesses from the manufacturer. The scattering loss \cite{SuzanneBvanDam.2018}
\begin{equation}
 \mathcal{L}_{\mathrm{scat}}=   \sin^2 \left( \frac{2 \pi n_{\mathrm{d}}t_{\mathrm{d}}}{\lambda_0} \right) \cdot \frac{(1+n_{\mathrm{d}})(1-n_{\mathrm{d}})^2}{n_{\mathrm{d}}} \cdot \left(\frac{4 \pi \sigma_{\mathrm{rms}}}{\lambda_0}\right)^2,
\end{equation}
originates from the diamond surface roughness $\sigma_{\mathrm{rms}}$. Scattering can also occur on the diamond-mirror interface, depending on the termination of the mirror coating. Here, the mirror is terminated with a layer of high refractive index, leading to a field node at the surface such that scattering can be omitted. In addition, we include absorption loss $\mathcal{L}_{\mathrm{abs}} = 2\alpha_{\mathrm{d}} t_{\mathrm{d}}$ described by the absorption coefficient $\alpha_{\mathrm{d}}$ of the diamond sample. For the general grade diamond sample, which contains an increased nitrogen concentration $>100\,\mathrm{ppb}$, a broadband absorption with $\alpha_{\mathrm{d}}\sim 0.1 - 0.5$~cm$^{-1}$ is expected \cite{Friel2010}.

For all loss contributions, knowledge of the diamond thickness and the local mode character are important. Therefore, we evaluate a measurement of the cavity mode dispersion at one location, i.e., the mode frequency shift as a function of the mirror separation, and fit a simulation to the measured dispersion to obtain the thickness $t_{\mathrm{d},0} = 5.96 \, \upmu\mathrm{m}$. For the other locations along the line, we make use of the known mode composition change from diamond-like (D) to air-like (A) between a local finesse minimum and maximum, corresponding to a thickness change of $\lambda / \left(4n_{\mathrm{d}}\right) \approx 66 \, \mathrm{nm}$. We interpolate the membrane thickness along the line by a linear increase between consecutive extrema. 

We then perform a fit \cite{LMFIT} based on the described model and the interpolated thicknesses for the finesse data along the blue line in figure \ref{Fig3} (e) with the surface roughness $\sigma_{\mathrm{rms}}$ and the absorption coefficient $\alpha_{\mathrm{d}}$ as free parameters. The result is plotted in figure \ref{Fig3} (f) together with the measured data. We note that the deviation between fit and data around pixel 75 originates from a local defect present on the membrane, which can be seen at the respective position in figure \ref{Fig3} (e). From the fit we obtain a roughness of $\sigma_{\mathrm{rms}} = (1.02 \pm 0.03) \, \mathrm{nm}$ and an absorption coefficient of $\alpha_{\mathrm{d}} = (0.19 \pm 0.01) \, \mathrm{cm}^{-1}$. While the absorption coefficient lies within the expected range, the predicted surface roughness is higher than the measured values of $\sigma_{\mathrm{rms}} \approx 0.4 - 0.5 \, \mathrm{nm}$ \cite{Heupel.2020}. Therefore, we conclude that we still miss further loss contributions. Possible origins can be surface absorption due to sp$^2$ carbon and surface defect states, and loss originating from the mismatch between the curved cavity mode phase front and the planar membrane surface (see below). We note that by adding an anti-reflection coating on the membrane, one could separate loss originating from roughness and surface absorption.

\subsection*{Increased cavity losses due to transverse mode mixing} \label{sec:modemixing}
\begin{figure*}
    \includegraphics[width=0.98\textwidth]{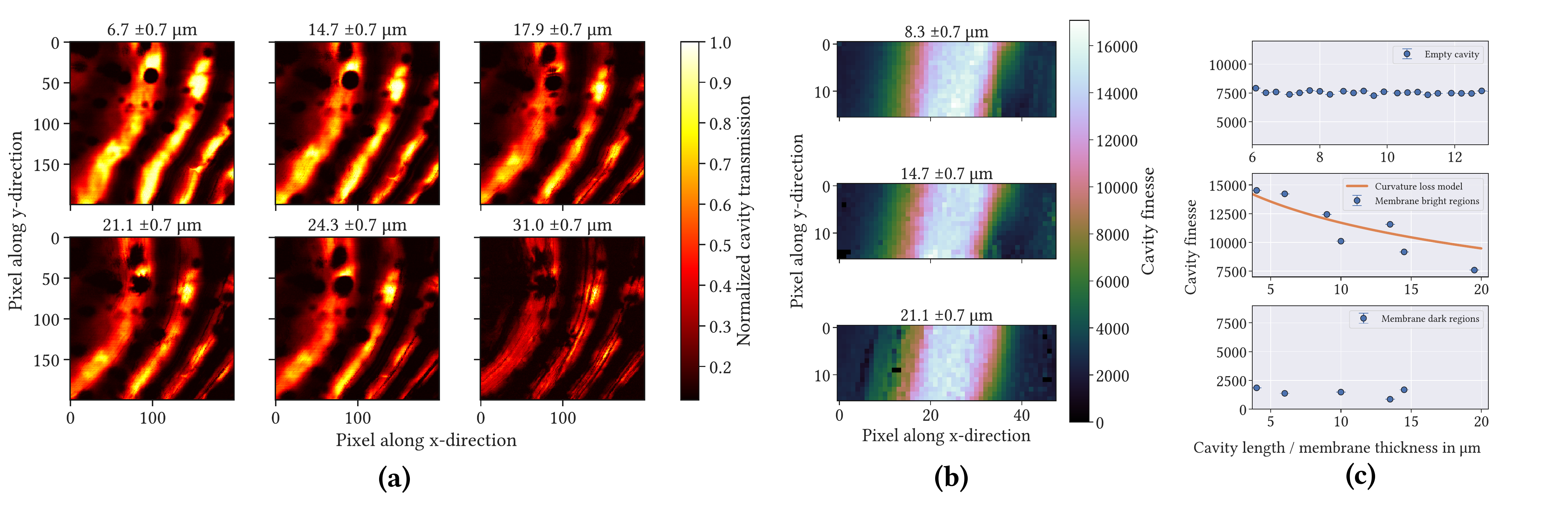}
\caption[]{\textbf{(a)} Cavity transmission scans for increasing air gap and fixed membrane thickness. The effective cavity length is given above each scan. The scans are normalized to the maximum value of all scans. \textbf{(b)} Cavity finesse scans for different effective cavity length by changing the air-gap on an air-like region. \textbf{(c)} Cavity finesse of the empty cavity (upper panel) for different cavity length and finesse on bright regions (central panel) and dark regions (lower panel) on the membrane for different membrane thicknesses. Each data point shows the average of at least 100 measurements and the orange curve in the central panel shows a model based on additional losses for air-like modes coming from the mode curvature at the diamond-air interface. For easier comparison, the range of shown cavity lengths in the upper panel corresponds to the effective cavity lengths resulting from the membrane thickness and the air gap of the lower two panels.}
\label{Fig4}
\end{figure*}

\begin{figure*}
    \includegraphics[width=0.8\textwidth]{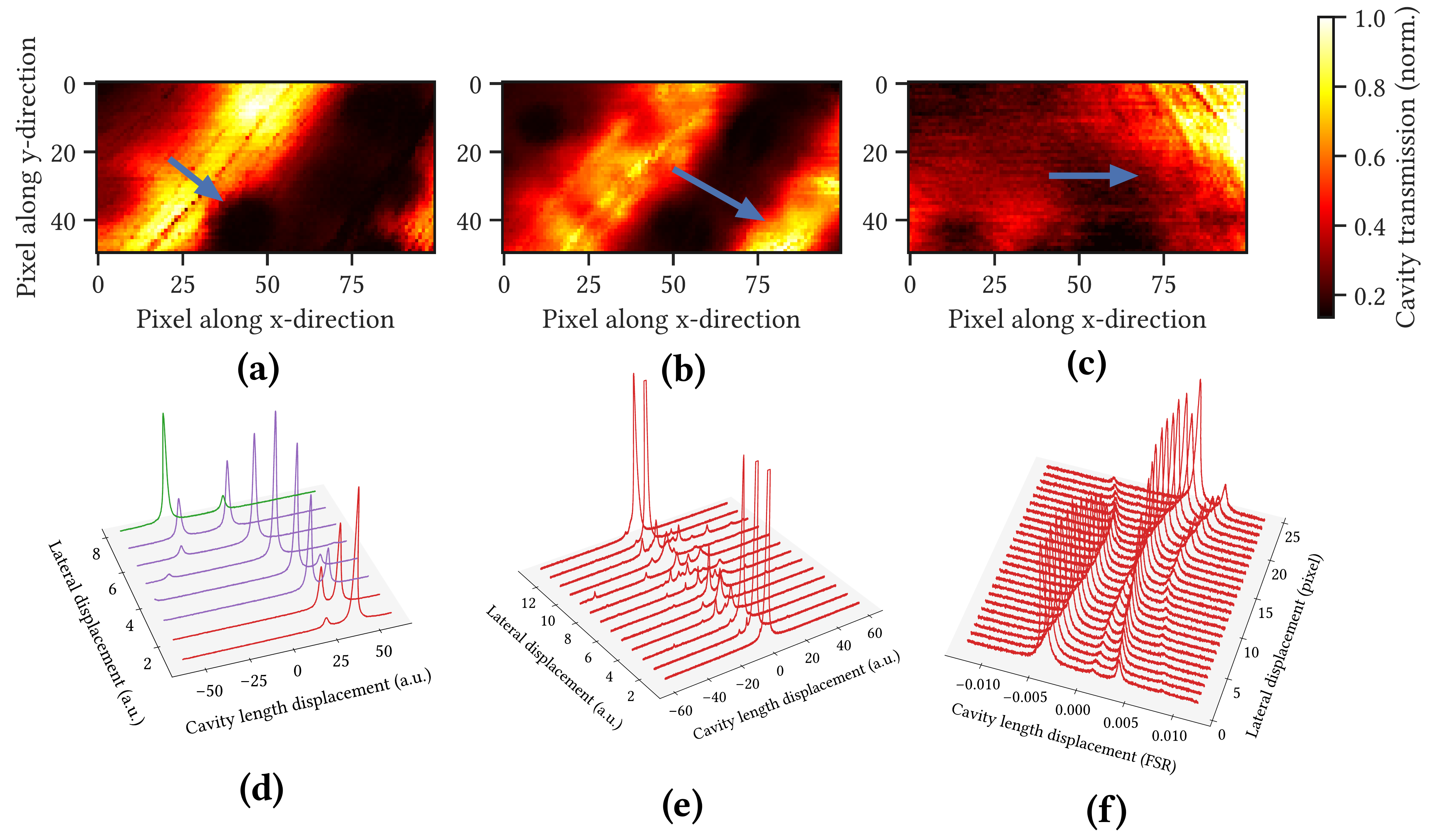}
\caption[]{\textbf{(a)}-\textbf{(c)} Cavity transmission scans of areas with resonant mode coupling. Blue arrows indicate the location and direction of the scans shown in \textbf{(d)} - \textbf{(f)}. \textbf{(d)} Cavity transmission spectra under air-like conditions for positions along the blue arrow in \textbf{(a)}. The color is changed after each avoided crossing for better visibility. \textbf{(e)} Transmission spectra under diamond-like conditions along the arrow shown in \textbf{(b)}. Coupling between the fundamental mode and a series of higher-order modes leads to a strong decrease of the cavity transmission. \textbf{(f)} Cavity transmission spectra for positions along the arrow shown in \textbf{(c)}.}
\label{Fig5}
\end{figure*}

In most spatial maps of the diamond membrane, we observe sharp lines of constant mode character where the cavity finesse and transmission are significantly lower than in the surrounding. Similar but less severe behavior has been observed also for bare mirrors, and resonant transverse-mode mixing has been identified as the origin \cite{Benedikter.2015,Benedikter.2019}. It occurs when higher-order transverse modes of a neighboring longitudinal mode order $q-1$ become resonant with a fundamental cavity mode of order $q$, and when boundaries such as the mirror surfaces do not match the phase fronts of Hermite-Gaussian modes, such that scattering leads to a coupling and hybridization of (near-) resonant modes. The effect is strongly dependent on the cavity length, and at large mirror separation, where low-order transverse modes fulfill the mode-mixing resonance condition, it becomes more dominant. To analyze the effect in the presence of a diamond membrane, we record transmission maps of one area (similar to figure \ref{Fig3} (a)) and change the air gap of the cavity over a large range. Six different scans are shown in figure \ref{Fig4} (a). As we increase the effective cavity length, we observe an overall transmission decrease as well as the sharp mode-mixing lines which become increasingly prominent. Notably, also at shortest mirror separation, we always observe mode-mixing lines in diamond-like regions, while they only appear at large distance in air-like regions.

To see whether the overall decrease in transmission with increasing mirror separation correlates with the cavity finesse, we perform finesse scans on an air-like sub-region for three different cavity lengths, see figure \ref{Fig4} (b). We observe no significant drop of the maximum finesse for the different cavity lengths, and only a weak sign of mode mixing within the studied cavity length range. This indicates that the transmission drop is dominated by mode matching.

In contrast, we have observed that losses under air-like conditions increase for thicker diamond membranes. To study this effect, we use a second diamond sample of electronic grade quality which features several membranes with different thicknesses and that was bonded on a second cavity mirror M$_{\mathrm{B}}$ (see \cite{Heupel.2020} for details on the sample). We perform cavity transmission scans on regions of different thicknesses between $6  \, \mathrm{\upmu m}$ and $19.5  \, \mathrm{\upmu m}$ and measure the cavity finesse on bright (high air-like character) and dark (high diamond-like character) regions. As depicted in the central panel of figure \ref{Fig4} (c), the cavity finesse on bright regions shows a strong decrease with increasing membrane thickness, dropping significantly faster than expected for bulk absorption. At the same time, we observe no drop of the cavity finesse on dark regions (diamond-like conditions) with increasing membrane thickness, shown in the lower panel of figure \ref{Fig4} (c). We note that for the investigated effective cavity lengths, the empty cavity shows a stable finesse (see figure \ref{Fig4} (c), upper panel) which is valid even for longer cavity lengths (for details see Supplementary Material \cite{supp}), thus ruling out effects related to the cavity fiber, i.e. clipping losses, as the origin of our observation. A plausible explanation is the shape mismatch of the diamond-air interface with the curved wavefront of the cavity mode \cite{Janitz.2015}, which becomes more prominent with increasing membrane thickness. There, the membrane thickness approaches the Rayleigh range of the cavity mode, and the curvature of the phase front at the diamond surface increases. Under such conditions, the cavity mode can no longer fulfill air-like conditions over the entire mode cross section, and an electric field at the interface will be present at the outer parts of the mode. This effectively reduces the air-like character of the mode and thus leads to increased scattering loss and mode mixing. As shown in the central panel of figure \ref{Fig4} (c), we can model the behavior with a simple estimate where we calculate the diamond / air character based on the weighted fraction of electric field present at the diamond-air interface due to the wavefront curvature for a given cavity and membrane geometry (see Supplementary Material \cite{supp}). Since the occurrence of mode-curvature-related effects scales with the ratio of the radius of curvature of the fiber mirror to the diamond thickness rather than with just the thickness, these effects can also be relevant for experiments with thinner membranes. Consistent with this picture, we observe no significant change of the cavity finesse under diamond-like conditions for increasing membrane thickness, as the overall losses are much higher and thus the relative change smaller.

Beyond this general trend which is present on the entire membrane and for any air gap, our measurements show that the diamond leads to significant resonant coupling between different transverse modes. To directly reveal the mode coupling, we use the cavity with fiber mirror F$_{\mathrm{B}}$ and the general grade sample to scan laterally in fine steps over a region where the transmission drops sharply and record the cavity modes at each position. We show transmission maps on three different areas where mode coupling occurs in figure \ref{Fig5} (a,b,c). To display the coupling of higher-order modes to the fundamental mode, we record cavity transmission spectra taken by varying the cavity length, for each position along three different linear paths indicated in figure  \ref{Fig5} (a,b,c), see figure \ref{Fig5} (d,e,f), respectively. At locations where a fundamental mode and a higher-order mode would show degeneracy, avoided crossings appear, and the transmission drops. Figure \ref{Fig5} (d) shows the cavity spectra for a scan across an air-like region. The first spectrum (red, bottom) features one prominent fundamental mode and a second, higher-order mode approaching from the left. When both modes would become degenerate, the coupling leads to an avoided crossing. In this example, a second avoided crossing with another transverse mode and larger coupling is visible at small spatial distance.

Under diamond-like conditions, mode mixing is observed to be more severe, and it can include a large number of consecutive avoided crossings such that the transmission drops almost completely, see figure \ref{Fig5} (e). Here, the laser power is increased to saturate the APD outside the avoided crossings such that the modes can still be detected while mixing. A further example with a large spatial region of mode mixing is shown in figure \ref{Fig5} (f). Here, one can observe the coupling of both polarization modes. For calibration of the cavity length changes, each spectrum is recorded over more than one FSR.

The alignment of mode mixing resonance contours with the diamond membrane topography can be understood by the different effect of the membrane on the frequency of different transverse modes \cite{Sankey2010}. Due to the larger Gouy phase of higher-order modes compared to the fundamental mode, their diamond- and air-like mode character occur for different membrane thicknesses, such that thickness variations will lead to a large variation of mode frequency differences, and transverse-mode resonance conditions will follow isocontours of constant membrane thickness.

Several contributions add to the mode mixing, such as imperfect mirror shape and misalignment of the cavity (see e.g. fig. \ref{Fig3} (d) vs (e)), which can be both minimized. The origin of the membrane-induced mode mixing is due to imperfect surface topography such as a wedge, but also due to the shape mismatch between the planar diamond-air interface and the parabolic wave front of the cavity mode \cite{Janitz.2015}. Due to this curvature, an exact air-like condition with zero field at the entire interface cannot be achieved. This introduces interface-induced mode mixing also for air-like modes. For a plano-concave cavity geometry, the wave front curvature approaches a planar shape close to the waist on the plane mirror, such that for thin membranes, one can largely avoid diamond-induced mode mixing \cite{Haeussler2019,Flagan2021}.  

\subsection*{Polarization mode splitting and birefringence} \label{sec:polarization}
\begin{figure}
    \includegraphics[width=0.5\textwidth]{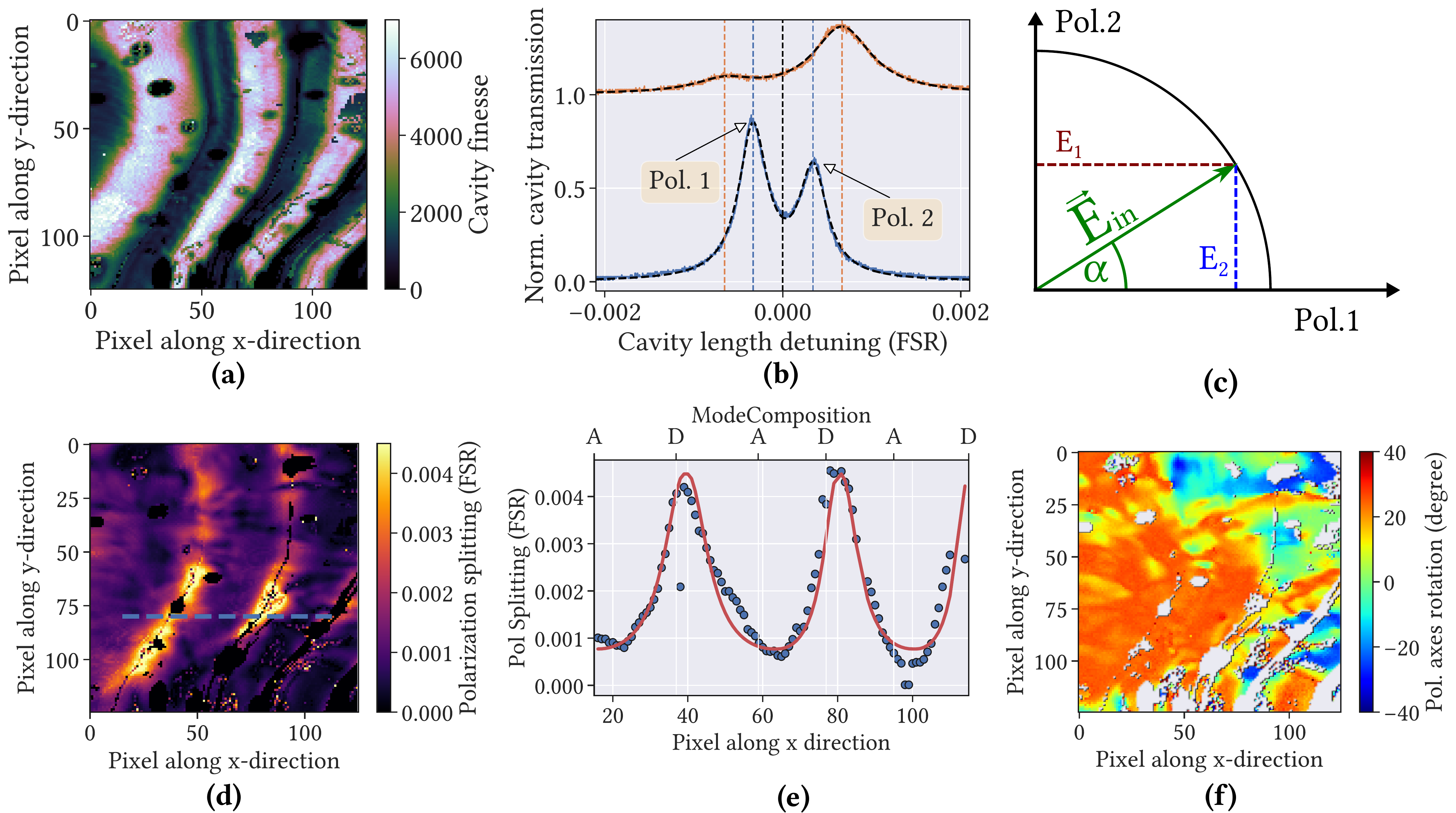}
\caption[]{Polarization splitting and rotation of the fundamental cavity modes. \textbf{(a)} Cavity finesse scan on the diamond membrane. \textbf{(b)} Fundamental cavity modes on a position with high (blue) and low (orange) finesse. The data for low finesse is shifted along the y-axis for better visibility. Black dashed lines are Lorentzian fits on the transmission data. Vertical lines show the different amount of mode splitting. \textbf{(c)} The intensity ratio of the polarization modes can be used to determine the relative polarization angle between incoupled light and the cavity polarization modes. \textbf{(d)} Lateral scan of the polarization splitting in units of FSR. One finds a larger splitting for modes with higher diamond-like character. \textbf{(e)} Line-cut along the dashed blue line in (d). The red line shows a fit. \textbf{(f)} Spatial map of the polarization angle obtained from the same data used for \textbf{(d)}.}
\label{Fig6}
\end{figure}

Finally, we investigate how the diamond membrane affects the polarization modes. For cavity mirrors with broken rotational symmetry, an intrinsic polarization mode splitting is present \cite{Uphoff_2015}, which is also the case for the cavities used here. For the empty cavity, we measure a mode splitting of $\delta=\nu_\mathrm{split}/\nu_\mathrm{FSR}=3.82\times10^{-5}$, which is consistent with the measured ellipticity of the concave profile of the fiber (see Supplementary Material \cite{supp}). On the diamond membrane, we observe a significantly increased and spatially varying mode splitting spanning values between $\delta=5\times10^{-4}$ and $5\times10^{-3}$. Figure \ref{Fig6} (a) shows a finesse scan of the investigated region of the membrane for reference of the mode character. From cavity transmission spectra on two locations with different mode character, we find a significant difference in the polarization splitting, see figure \ref{Fig6} (b). When measuring a map of the polarization splitting on the same area as shown in figure \ref{Fig6} (a), we observe fringes with the same shape as for the cavity finesse, see figure \ref{Fig6} (d). Positions with low finesse, i.e. diamond-like character, show a large splitting, and positions with air-like character show a smaller splitting. 
We suggest the mode-character dependent energy distribution as an origin of this variation in the splitting and model it with the relative intensity in the diamond that was already used for the effective losses as
\begin{equation}
    \delta = \frac{n_{\mathrm{d}}E_{\mathrm{max,d}}^2}{E_{\mathrm{max,a}}^2} \cdot \delta_0.
\end{equation}
A fit of this model to a line of the data from figure \ref{Fig6} (d) is shown in figure \ref{Fig6} (e) and yields $\delta_0 = (1.86 \pm 0.04) \times 10^{-3}$. Despite a few outliers and a non-perfect line shape, the fit confirms the mode character as a major contribution to the variation of the polarisation splitting.

Even the smallest observed values for the polarization splitting within regions with high air-like character are increased compared to the empty cavity. We attribute this to birefringence of the diamond which originates from local strain. The observed diamond-induced mode splitting in air-like regions of $\delta=0.001$ corresponds to a refractive index difference of $\Delta n=\lambda/t_\mathrm{d} \cdot \delta/2 =5\times 10^{-5}$, indicating an elevated local strain level.

Further, we observe a spatial variation of the intensity ratio of the polarization modes, originating from a rotation of the cavity polarization axes. To study this observation in more detail, we measure the intensity ratio of both polarization modes and calculate the corresponding polarization angle between the probe laser and one of the cavity modes by $\tan(\alpha) = E_2 / E_1 = \sqrt{I_2/I_1}$, as sketched in figure \ref{Fig6} (c). Here $I_1$ and $I_2$ are the transmitted intensities of the two polarization eigenmodes. As shown in figure \ref{Fig6} (f), the extracted angle changes significantly over the region, showing a completely different pattern than the topography of the membrane. We ascribe this signal to the variation of the orientation of the slow axis of the birefringent diamond, which is consistent with a spatially varying strain. To exclude measurement artefacts, we have repeated the measurements with a second cavity with the fiber mirror F$_{\mathrm{B}}$ at a slightly shifted position (see Supplementary Material \cite{supp}), and observe the same pattern.
For comparison, we also studied a second electronic grade diamond sample in a cavity. There, we observed mode-character dependent polarization splitting between $\delta=2\times10^{-4}$ and $9\times10^{-4}$, resulting in a reduced birefringence of $\Delta n\approx 1.6 \times10^{-5}$ (see Supplementary Material \cite{supp}). This lower level of birefringence together with the splitting induced by the ellipticity of the fiber mirror can lead to a polarization mode splitting that remains smaller than a cavity linewidth as long as the finesse remains smaller than $\approx 5000$. We note that we have observed also electronic grade samples with larger birefringence. This emphasizes the need to carefully select optimal samples, where one to two orders of magnitude smaller strain levels have been reported \cite{Friel2009}.

It remains an interesting question whether noticeable strain can originate from the van der Waals bond, e.g. for non-perfectly planar interfaces. This contribution could be quantified by measuring the birefringence before and after bonding. Given that the nitrogen vacancy center in diamond is sensitive to local strain for both the optical as well as the spin levels \cite{Batalov_2008,Udvarhelyi_2018}, this might be checked in following studies by the properties of NV centers in addition to the cavity properties. Birefringence can become a limiting factor for aligning the cavity mode polarization with the transition dipole orientation of color centers. This can reduce the achievable coupling strength, lead to issues with circularly polarized transitions, and limit the performance of the crossed polarization detection scheme \cite{Ruf2021}. On the other hand, suitable polarization mode splitting can be made use of for polarization-selective Purcell enhancement for efficient resonance fluorescence collection \cite{Wang2019,Tomm2021}.

\section*{Conclusions} \label{sec:conclusions}
In our study we have used scanning cavity microscopy to evidence the effects of the topography of a diamond membrane on the cavity modes. The diamond- and air-like character strongly dominates the mode properties and cavity loss, and we were able to differentiate the different loss contributions and show the importance of absorption loss for samples with increased nitrogen concentration. We have further identified transverse mode mixing as a limiting mechanism when operating under diamond-like conditions, and increasing loss of air-like modes when the membrane thickness becomes comparable to the mirror radius of curvature. 
Finally, we have introduced a technique to sensitively measure local birefringence and observed a mode-character dependent polarization mode splitting.
The observed loss and birefringence effects become smaller for thinner membranes, such that one can target a compromise between improved emitter coherence for thicker, and improved cavity performance for thinner membranes. Furthermore, curvature-related cavity loss at the membrane-air interface as well as mode mixing depend on the ratio between the membrane thickness and the mirror radius of curvature, suggesting an optimal choice for the radius of curvature. 
Our results thereby provide insight into the required cavity geometry and the properties of membranes for optimized spin-photon interfaces.

\section*{Acknowledgements} \label{sec:acknowledgements}
We acknowledge experimental support from Jonas Grammel, Timon Eichhorn, and Tobias Krom. This project received funding from the European Union Horizon 2020 research and innovation program within the Quantum Flagship project SQUARE (Grant Agreement No. 820391), the German Federal Ministry of Education and Research (Bundesministerium für Bildung und Forschung, BMBF) within the project Q.Link.X (Contracts No. 16KIS0879 and 16KIS0877), QR.X (Contracts No. 16KISQ004 and 16KISQ005), SPINNING (Contract No. 13N16211), and NEQSIS (Contract No. 16KISQ029K), and the Karlsruhe School of Optics and Photonics (KSOP).

\end{document}

\title{Supplementary material: Scanning cavity microscopy of a single-crystal diamond membrane integrated into a fiber-based Fabry-Pérot cavity}
\onecolumngrid

\author{Jonathan Körber}
\affiliation{ 
Physikalisches Institut, Karlsruhe Institute of Technology (KIT), Wolfgang-Gaede Str. 1, 76131 Karlsruhe, Germany}

\author{Maximilian Pallmann}
\affiliation{ 
Physikalisches Institut, Karlsruhe Institute of Technology (KIT), Wolfgang-Gaede Str. 1, 76131 Karlsruhe, Germany}

\author{Julia Heupel}
\affiliation{Institute of Nanostructure Technologies and Analytics (INA), Center for Interdisciplinary Nanostructure Science and Technology (CINSaT), University of Kassel, Heinrich-Plett-Straße 40, 34132 Kassel, Germany}

\author{Rainer Stöhr}
\affiliation{3rd Institute of Physics, University of Stuttgart, Pfaffenwaldring 57, 70569 Stuttgart, Germany}

\author{Evgenij Vasilenko}
\affiliation{ 
Physikalisches Institut, Karlsruhe Institute of Technology (KIT), Wolfgang-Gaede Str. 1, 76131 Karlsruhe, Germany}
\affiliation{Institute for Quantum Materials and Technologies (IQMT), Karlsruhe Institute of Technology (KIT), Herrmann-von-Helmholtz Platz 1, 76344 Eggenstein-Leopoldshafen, Germany}

\author{Thomas Hümmer}
\affiliation{Faculty of Physics, Ludwig-Maximilians-University (LMU), Schellingstr. 4, 80799 Munich, Germany}

\author{Larissa Kohler}
\affiliation{ 
Physikalisches Institut, Karlsruhe Institute of Technology (KIT), Wolfgang-Gaede Str. 1, 76131 Karlsruhe, Germany}

\author{Cyril Popov}
\affiliation{Institute of Nanostructure Technologies and Analytics (INA), Center for Interdisciplinary Nanostructure Science and Technology (CINSaT), University of Kassel, Heinrich-Plett-Straße 40, 34132 Kassel, Germany}

\author{David Hunger}
\affiliation{ 
Physikalisches Institut, Karlsruhe Institute of Technology (KIT), Wolfgang-Gaede Str. 1, 76131 Karlsruhe, Germany}
\affiliation{Institute for Quantum Materials and Technologies (IQMT), Karlsruhe Institute of Technology (KIT), Herrmann-von-Helmholtz Platz 1, 76344 Eggenstein-Leopoldshafen, Germany}
\email{david.hunger@kit.edu}

\date{\today} 

\maketitle

\section{Characterization of the fiber mirrors}
\label{Sec:FiberDetails}
\begin{figure}[tb]
    \centering
    \includegraphics[width=0.7\textwidth]{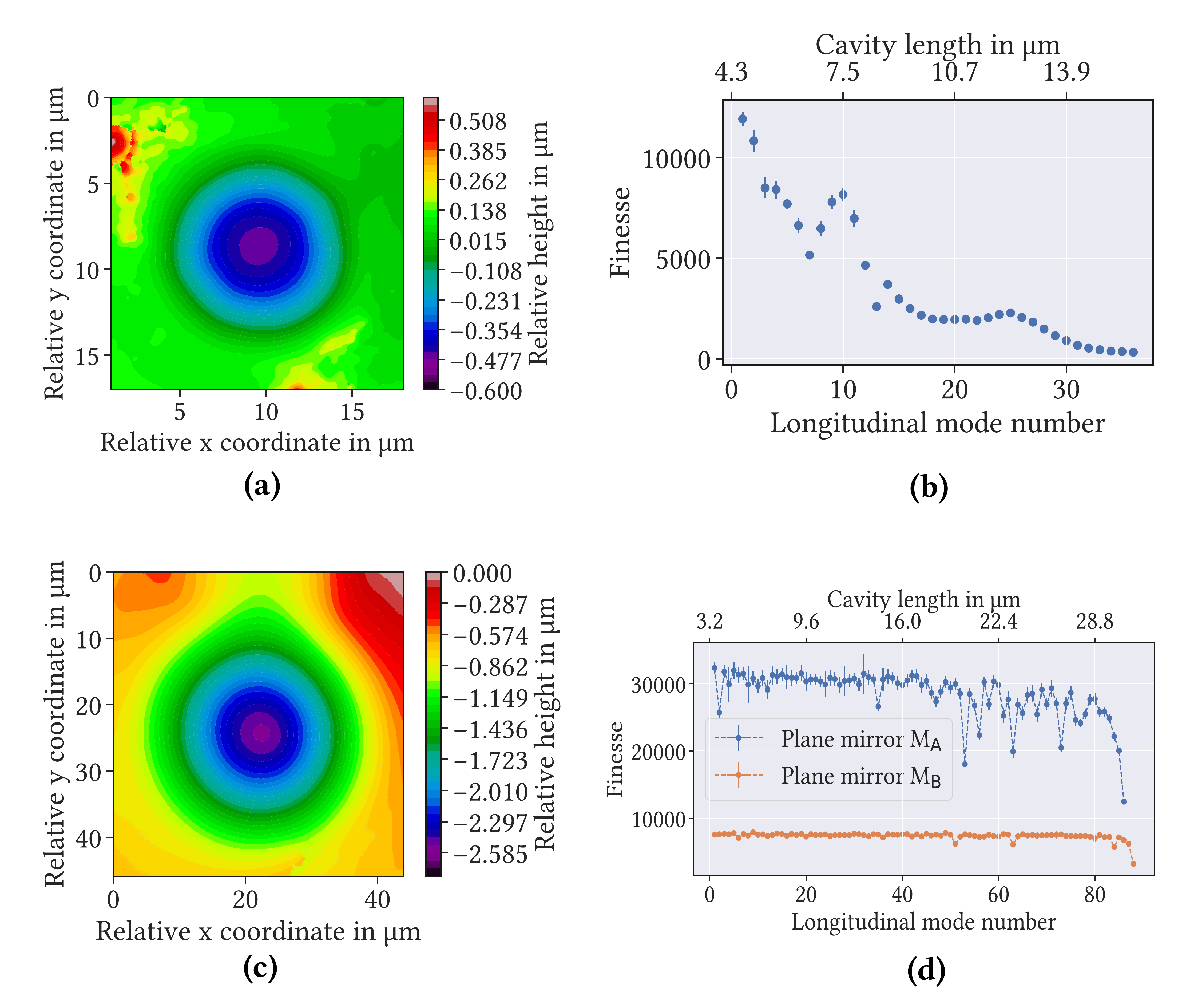}
\caption[]{Height profiles and performance of the fiber mirrors \textbf{(a)} WLI-measured height profile of fiber mirror F$_{\mathrm{A}}$ (tilt-corrected). \textbf{(b)} Measured finesse for different longitudinal mode order and a cavity made from F$_{\mathrm{A}}$ and M$_{\mathrm{A}}$. \textbf{(c)} WLI-measured height profile of fiber mirror F$_{\mathrm{B}}$ (tilt-corrected). \textbf{(d)} Measured finesse for different longitudinal mode order and a cavity made from F$_{\mathrm{B}}$ and M$_{\mathrm{A}}$ (blue) and additionally from F$_{\mathrm{B}}$ and M$_{\mathrm{B}}$ (orange). }
\label{Fig1}
\end{figure}
For characterization of the used fiber mirrors F$_{\mathrm{A}}$ and F$_{\mathrm{B}}$ respectively we use a home-built WLI to reconstruct the concave profiles on the end facets after the coating process. Therefore, we use a commercial Mirau objective (Nikon: CF IC Epi, 50x, 0.55 NA), mount the fibers (or the mirror with membrane) on a 3-axis micropositioning stage (PI Micos, 200nm positioning reproducibility), and perform a 5-step reconstruction using an objective scanner (nPoint) with 1nm closed loop positioning accuracy. The resulting height profiles are shown in figure \ref{Fig1} (a) and (c). Fiber mirror F$_{\mathrm{A}}$ is characterized by a total profile diameter of approximately $10 \, \upmu \mathrm{m}$ and a depth of $0.6 \,\upmu \mathrm{m}$. Fiber mirror F$_{\mathrm{B}}$ shows a larger diameter of $30 \, \upmu \mathrm{m}$ and a depth slightly larger than $2 \, \upmu \mathrm{m}$. Furthermore, we fit the profiles over a region of $4 \, \upmu \mathrm{m}$ around the center with elliptical 2D-parabolic profiles to infer the ROCs across the two half axes for each profile. We estimate ROCs of $33.1 \, \upmu \mathrm{m}$ and $30.5 \, \upmu \mathrm{m}$ for F$_{\mathrm{A}}$ as well as $55.2 \, \upmu \mathrm{m}$ and $48.7 \, \upmu \mathrm{m}$ for F$_{\mathrm{B}}$.

To check the cavity performance of both fiber mirrors together with the plane mirror we measure the finesse as described in the section \textit{Methods and materials} of the main manuscript for different cavity lengths after careful angular alignment of the fiber mirror in front of the plane mirror, using our TDL at $\lambda = 639.7 \, \mathrm{nm}$. In addition to the plane mirror M$_{\mathrm{A}}$ that is used for the measurements of the main manuscript we investigate a second mirror M$_{\mathrm{B}}$ that is used to repeat some of the measurements from the main manuscript with a different diamond sample. While plane mirror M$_{\mathrm{A}}$ has a high-index termination, mirror M$_{\mathrm{B}}$ is terminated with a low-index spacer layer since it was designed for the use with shallowly implanted emitters. For calibration of the physical distance between the fiber and the plane mirror at the lowest accessible longitudinal mode number (labeled as 1) before physical contact, we measure cavity resonances of a broadband light source at a fixed cavity length in transmission and calculate the distance using
\begin{equation}
    L_{\mathrm{cav}} = \frac{\lambda_1 \cdot \lambda_2}{2 \cdot \left( \lambda_2 - \lambda_1 \right)}
\end{equation}
from two consecutive resonances of increasing wavelength $\lambda_1$ and $\lambda_2$. \\
As shown in figure \ref{Fig1} (b), the finesse of the cavity made from fiber mirror F$_{\mathrm{A}}$ and plane mirror M$_{\mathrm{A}}$ strongly drops with increasing cavity length, starting from the lowest accessible cavity length of $L^{\mathrm{min}}_{\mathrm{cav1}} = 4.64 \, \upmu \mathrm{m}$ and breaks down completely for lengths larger than $14 \, \upmu \mathrm{m}$. The maximum measured finesse of $\mathcal{F}^{\mathrm{cav1}}_{\mathrm{max}} = 11915$ at the lowest achievable cavity length already shows additional losses of $418 \, \mathrm{ppm}$ to the estimated upper bound of finesse (calculated by the transmission of the coatings as only loss source) ${F}^{\mathrm{cav1}}_{\mathrm{ub}} = 57644$ that cannot be explained by absorption of the mirror coatings only and indicate together with the fast and strong decline of the cavity finesse systematic problems of the fiber mirror F$_{\mathrm{A}}$ (problems with the plane mirror M$_{\mathrm{A}}$ are excluded by its good performance together with the fiber mirror F$_{\mathrm{B}}$ as discussed later). We note that our results of the section \textit{Mode-character-dependent cavity loss} of the main manuscript hold despite these systematic issues, since the losses induced by the mode composition are higher than the systematic losses of the fiber mirror within the used cavity lengths. Also, we investigated the same effects using the second fiber mirror that does not show such high losses. The minimal accessible distance for F$_{\mathrm{A}}$ appears high with $L^{\mathrm{min}}_{\mathrm{cav1}} = 4.64 \, \upmu \mathrm{m}$, since its profile depth is below $1 \, \upmu \mathrm{m}$. We explain this by certain defects protruding the end facet of the fiber mirror that we investigated with the WLI, causing an early physical contact between fiber mirror and plane mirror. 

Finesse characterization of a cavity made from fiber mirror F$_{\mathrm{B}}$ and plane mirror M$_{\mathrm{A}}$ (plane mirror M$_{\mathrm{B}}$) are shown in figure \ref{Fig1} (d) in blue (orange). Overall, the measured cavity finesse is very stable with only small drops for particular mode numbers up to a cavity length of $29 \, \upmu \mathrm{m}$. In addition, the maximum measured finesse with plane mirror M$_{\mathrm{A}}$ (plane mirror M$_{\mathrm{B}}$) of $\mathcal{F}^{\mathrm{cav2}}_{\mathrm{max}} = 32381$ $\left(\mathcal{F}^{\mathrm{cav3}}_{\mathrm{max}} = 7912\right)$ are close to the estimated upper bound of the finesse values of $\mathcal{F}^{\mathrm{cav2}}_{\mathrm{ub}} = 55116$ $\left(\mathcal{F}^{\mathrm{cav3}}_{\mathrm{ub}} = 8025 \right)$ and show much lower additional losses of $80 \, \mathrm{ppm}$ $\left(11\, \mathrm{ppm}\right)$. Therefore, we use fiber mirror F$_{\mathrm{B}}$ for our quantitative measurements and for checking the cavity transmission with integrated membrane for different cavity lengths. 

\section{Lateral calibration of cavity scans}
\begin{figure}[tb]
    \centering
    \includegraphics[width=0.7\textwidth]{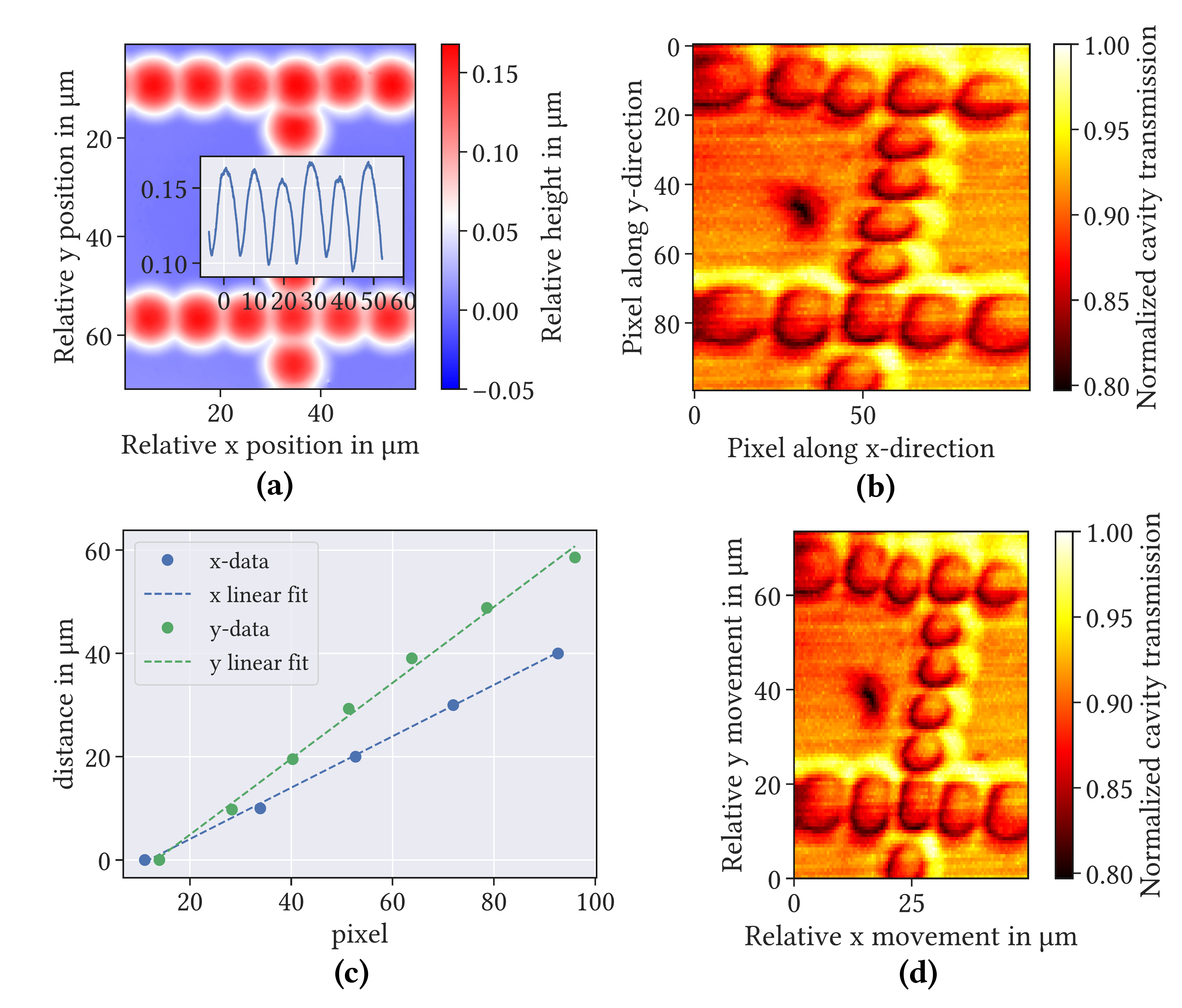}
\caption[]{Calibration of cavity transmission scans. \textbf{(a)} WLI-measurement of the periodical structure created on mirror M$_{\mathrm{B}}$ for calibration. The inset shows a cut through the upper horizontal row of bulges centered on the maximum of the first bulge to show that they are separated by roughly $10 \, \upmu\mathrm{m}$. \textbf{(b)} Non-calibrated cavity transmission scan of a region on  M$_{\mathrm{B}}$ showing the fabricated structure. \textbf{(c)} Distance between consecutive bulges over their pixel position within the transmission scan. Shown are the upper horizontal row (blue) and the vertical row (green). A calibration of both axes is done using a linear regression (dashed lines). \textbf{(d)} Cavity transmission scan after calibration of the axes (same data as \textbf{(a)}).}
\label{Fig2}
\end{figure}
For comparing our cavity transmission scans to WLI-height maps of the diamond sample we need to calibrate the lateral dimensions of our transmission scans. To this end, we use a pattern of equally separated bulges $(spacing = 10 \, \upmu \mathrm{m}$) that is produced by weak CO$_2$-laser pulses directly on the coating of a plane mirror (M$_{\mathrm{B}}$). A WLI-reconstructed height profile of a part of this fabricated structure is shown in figure \ref{Fig2} (a). The pattern is easily visible within cavity transmission scans as shown in figure \ref{Fig1} (b). As a consequence of not perfect placement of the mirror, the structure is slightly rotated. However, one can clearly see, that the features in the central part of the scan appear distorted. 

To calibrate the lateral pixels in the cavity scan to physical lengths we extract the pixel position of the centers of the bulges as well as the rotation angle. Subsequently, we plot the distance between consecutive bulges ($10 \, \upmu \mathrm{m}$) starting from zero along the x- and y-axis of the transmission scan corrected by estimated rotation angle over the pixel position of the respective bulge center and use a linear regression to estimate the length per pixel factor for each axis, respectively. Finally, we apply the extracted factors to obtain a calibrated transmission scan, as shown in figure \ref{Fig2} (d). Features in the center of the resulting scan now look much less distorted. We note that instead of using a simple linear regression one could also improve the calibration by using a more complex function (e.g. a cubic line) in the fit in figure \ref{Fig2} (c) which eventually describes the data better (especially in the outer parts). However, since we only need the calibration for a rough estimate, we use a linear regression for simplicity at transferring the calibration to the 2D-image data.

\section{Extraction of the hybridized mode composition}
\begin{figure}[tb]
    \centering
    \includegraphics[width=0.9\textwidth]{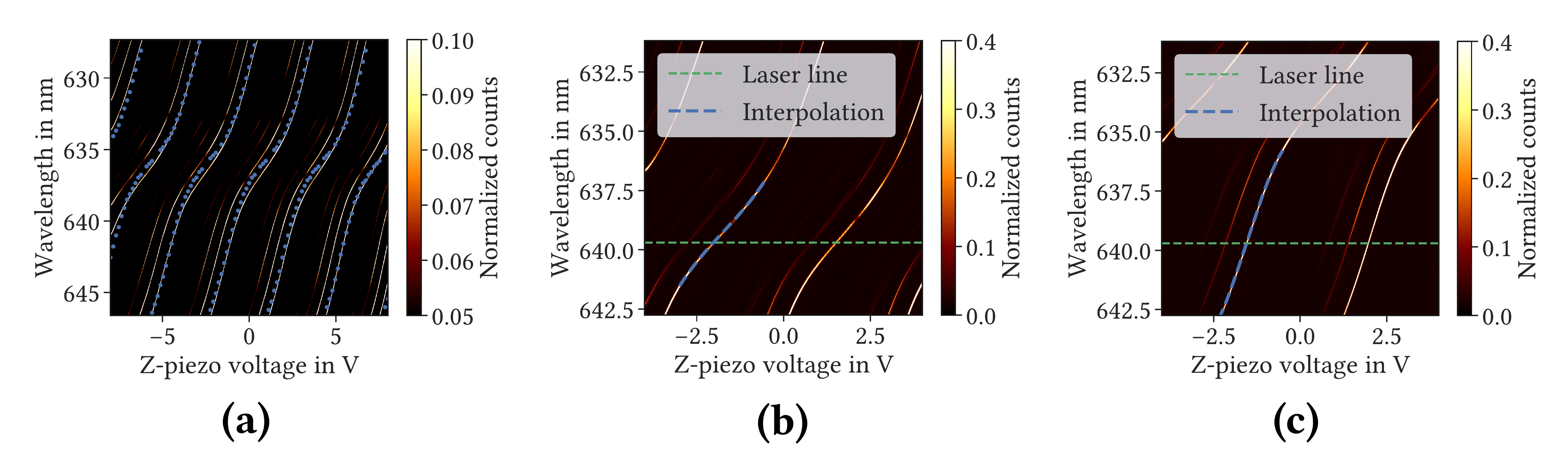}
\caption[]{Hybridized mode dispersion and extraction of mode composition. \textbf{(a)} Measurement of the hybridized mode dispersion at an arbitrary position on the membrane. The measured cavity transmission upon coupling a white light laser source is measured with the spectrometer and normalized. The color map is chosen such that also weaker signals (i.e. higher order modes) can be seen. Fundamental cavity modes show a flattening of the linear mode dispersion as a consequence of the mode hybridization. Blue points show a numerical simulation of the mode dispersion of consecutive fundamental cavity modes for a diamond thickness of $t_{\mathrm{d}} = 6.00 \, \upmu \mathrm{m}$ and an air gap of $6.75 - 5.20 \, \upmu \mathrm{m}$. \textbf{(b)} Mode dispersion on a position with high diamond-like character (pixel 34 in figure 3 of the main manuscript). To infer the slope of the dispersion the data of a fundamental mode around the laser line of $639.7 \, \mathrm{nm}$ (green dashed line) is interpolated by the blue dashed line. The slope of the interpolation at the laser line is used to estimate the mode composition. \textbf{(c)} Mode dispersion on a position with high air-like character (pixel 46 in figure 3 (a)-(c) of the main manuscript). A higher slope compared to \textbf{(b)} of the dispersion at the laser line indicates a steeper air-like character of fundamental modes at this position on the membrane.}
\label{Fig3}
\end{figure}
In order to infer the membrane-thickness dependent composition of fundamental cavity modes at particular positions on the membrane of our general grade sample we measure the mode dispersion by coupling a white light laser into the cavity and measuring transmission spectra as a function of the air gap (distance between the fiber mirror and plane mirror). The linear dependence between spectral cavity resonances and cavity air gap is modulated as a consequence of the hybridization of the modes due to the presence of the diamond membrane. This modulation can be seen clearly in the dispersion measurement for an arbitrary position on the membrane in figure \ref{Fig3} (a). A numerical simulation with a transfer matrix model - taking into account both DBR mirrors as well as the diamond membrane shows good agreement with the measurement for a diamond thickness of $t_{\mathrm{d}} = 6 \, \upmu \mathrm{m}$ and air gap values between $6.75 - 5.20 \, \upmu \mathrm{m}$. 

To extract the composition of fundamental modes at the spectral position of our characterization laser ($639.7 \, \mathrm{nm}$) we interpolate the measured mode dispersion and extract the slope of the dispersion for $639.7 \, \mathrm{nm}$. This is shown as an example for two measurements at a different membrane position (i.e. different membrane thickness) in figure \ref{Fig3} (b) and (c). The extracted absolute slopes are $1.61 \, \mathrm{nm/V}$ for figure \ref{Fig3} (b) and $4.47 \, \mathrm{nm/V}$ for figure \ref{Fig3} (c), respectively. As a consequence of the steeper slope, fundamental modes at the membrane position of figure \ref{Fig3} (c) have a much higher air-like character compared to modes at the position of figure \ref{Fig3} (b). Finally, we normalize the measured slopes of all investigated positions to the maximum measured slope and subtract them by the lowest, thus generating composition values between zero (high diamond-like character) and one (high air-like character). We note that this normalization is only meant for easier readability. Since we are only investigating the slope of particular points relative to each other we cannot claim pure air- or diamond-like modes from our measurements.

\section{Modeling of mode curvature causing additional loss}
\begin{figure}[tb]
    \centering
    \includegraphics[width=0.9\textwidth]{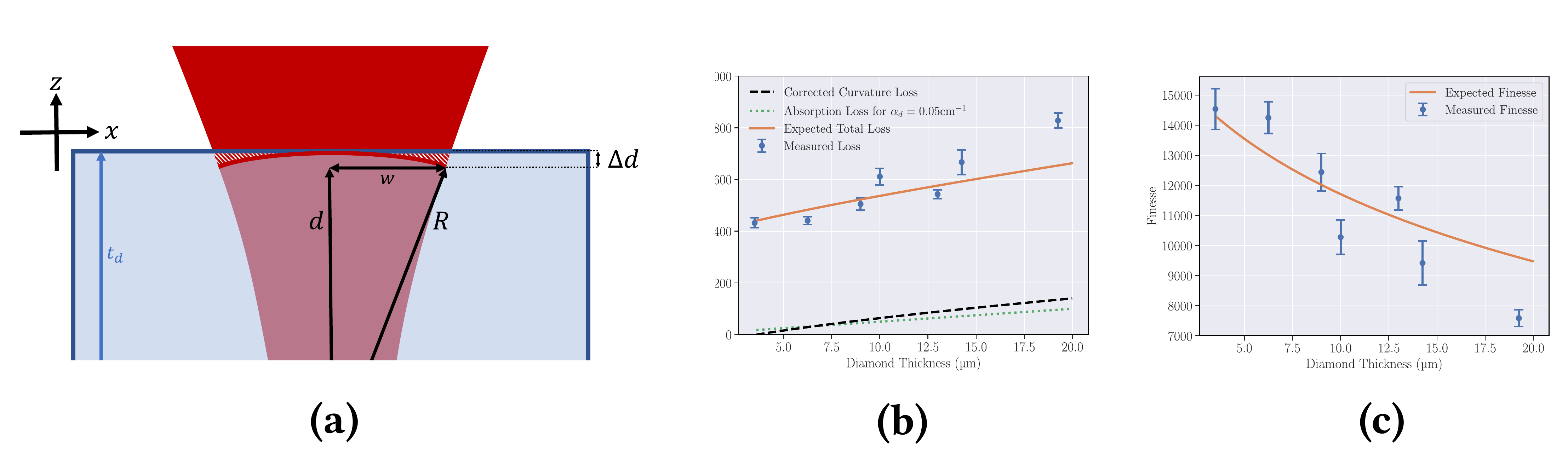}
\caption[]{Geometrical model for additional losses by leakage of diamond-like modes into the outer region of air-like modes. \textbf{(a)} Schematic of the diamond-air interface showing the essential parameters. The mode curvature leads to a mode volume where the effective diamond thickness is altered (red dashed area). \textbf{(b)} Estimated total loss, including loss by mirror transmission, absorption loss and the additional loss by mode curvature. \textbf{(c)} Respective expected Finesse, together with the measured values from fig. 4 (b) of the main paper.}
\label{Fig4}
\end{figure}

In order to gauge the additional loss that we measure for air-like modes with increasing diamond thickness as described in the section \textit{Increased cavity losses due to transverse mode
mixing} of the main paper, we developed a simple geometrical model of the mode curvature at the diamond-air interface. We start with a cavity mode that we assume to be perfectly air-like at the center of the mode, thus facing minimal scattering and transmission loss. The thickness difference between the center and the edge of the Gaussian mode is then given by
\begin{align}
    \Delta d=R-d=R-\sqrt{R^2-w^2}
\end{align}
where $R$ is the mode's radius of curvature, $d$ is the projection of $R$ onto the optical axis when pointing to the most outer mode region and $w$ is the beam radius at height $d$ (see fig.\ref{Fig4}(a)). For Gaussian beams, the radius of curvature and beam radius are given by
\begin{align}
    R(z)=z\cdot \left(1+\frac{\pi w_0^2 n_d}{\lambda z}\right)^2
\end{align}
\begin{align}
    w(z)=\left(w_0^2\cdot \left(1+\frac{\lambda z}{\pi w_0^2 n_d}\right)^2\right)^{1/2}
\end{align}
with $w_0$ being the beam waist, $z$ the distance to the mirror surface, $\lambda$ the wavelength and $n_d$ the refractive index of diamond.
We can then define a dimensionless factor $C_{a/d}$ that describes the mode character shift at a given lateral position x with respect to the center where $1$ is a full shift from air- to diamond-like and vice versa:
\begin{align}
    C_{a/d}(x)=\frac{4n_d \Delta d}{\lambda}\cdot \frac{x}{w}.
\end{align}
Here, we approximate the mode curvature as linear along x since $x,w \ll d,R$. As loss in our system scales with the field intensity present at the source of the loss, we have to add the mode's Gaussian intensity profile as an additional factor. Integrating over the whole 2D-plane then gives us the total amount of diamond-like character in an air-like mode,
\begin{align}
    \mathcal{L}_{curv}=\int_{0}^{\infty}C_{a/d}(x)\cdot \left(\frac{w_0}{w}\right)^2\cdot e^{-2x^2/w^2}\mathrm{d}x.
\end{align}
Note, that $\mathcal{L}_{curv}$ is not the actual loss, but only the fraction of losses originating from diamond-like modes leaking into an air-like mode. To scale this up to our experiment, we use the lowest loss that we measured for diamond-like modes, multiply this with $\mathcal{L}_{curv}$ and add it to the total loss. This combined with an estimated absorption loss and the loss via mirror transmission is in good agreement with the measurement, see fig.\ref{Fig4}(b) and fig.\ref{Fig4}(c) for the respective finesse.

\section{Polarization splitting of the empty cavity}
\begin{figure}[tb]
    \centering
    \includegraphics[width=0.9\textwidth]{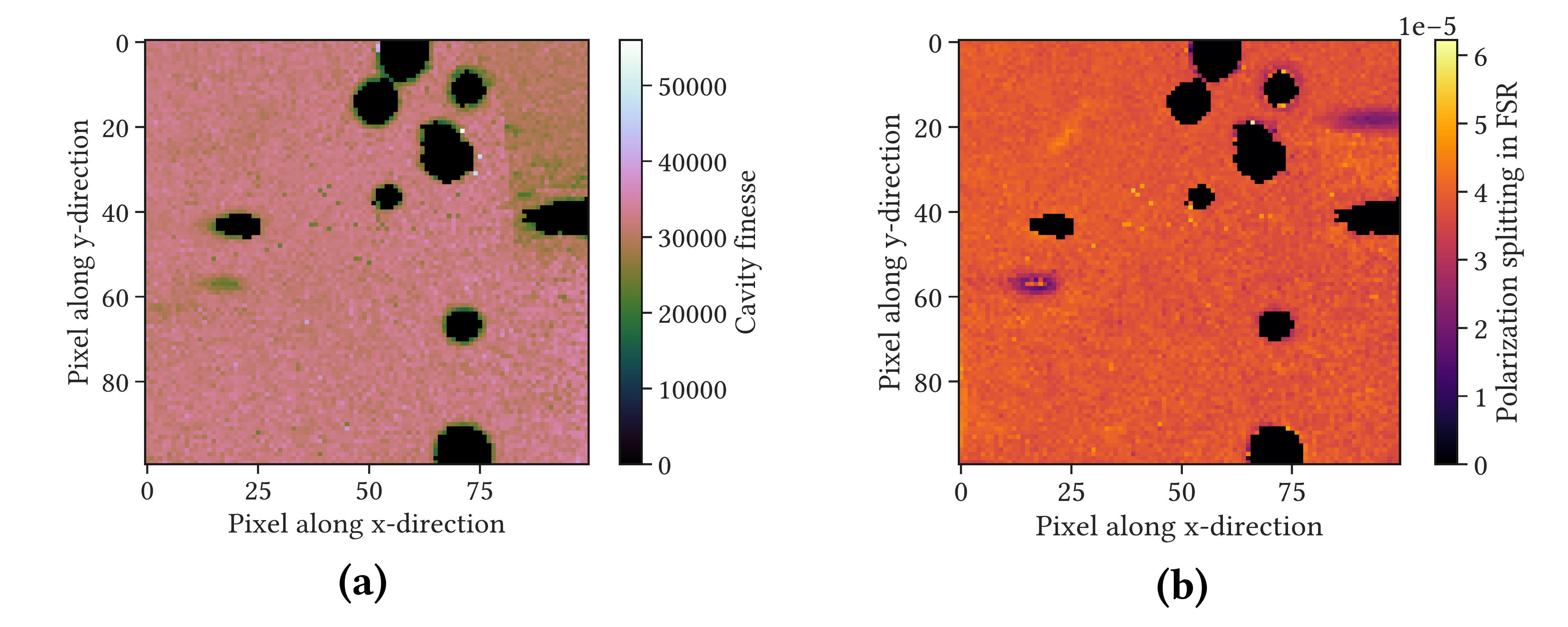}
\caption[]{Lateral cavity finesse and polarization splitting of fiber mirror F$_{\mathrm{B}}$ and plane mirror M$_{\mathrm{A}}$.  \textbf{(a)} Cavity finesse scan of fiber mirror F$_{\mathrm{B}}$ and plane mirror M$_{\mathrm{A}}$ without integrated diamond membrane. For each pixel the fiber was scanned at a rate of 20Hz and the finesse extracted as described in the section \textit{Methods and materials}. of the main manuscript. \textbf{(b)} Polarization splitting scan of fiber mirror F$_{\mathrm{B}}$ and plane mirror M$_{\mathrm{A}}$ without integrated diamond membrane. For each pixel the fiber was scanned at a rate of 20Hz and the splitting of two fundamental polarization modes was recorded.}
\label{Fig5}
\end{figure}

For separation of effects on the polarization splitting of fundamental cavity modes induced by the diamond membrane (as discussed in the section \textit{Polarization mode splitting and birefringence} of the main manuscript) from the effects resulting from the fiber mirror itself, we measured the polarization splitting properties of fiber mirror F$_{\mathrm{B}}$ in an empty cavity. As shown in figure \ref{Fig5} (a) the finesse of a cavity made from fiber mirror F$_{\mathrm{B}}$ and plane mirror M$_{\mathrm{A}}$ appears stable over a region of $100 \times 100 \, \upmu \mathrm{m}^2$ apart from isolated scatterers that we explain by dirt present on the mirror. We note that the average value of $31560 \pm 2175$ is very comparable to the values achieved when measuring a single position within the stable cavity length region in figure \ref{Fig1} (d). The consecutive scan of the polarization splitting in figure \ref{Fig5} yields an average value of $ \nu_{\mathrm{split}} / \nu_{\mathrm{FSR}} = \left(3.82 \pm 0.26 \right)\times10^{-5}$, that is used for comparison to the increased splitting with the present diamond membrane in \textit{Polarization mode splitting and birefringence} of the main manuscript.

To check the measurement with the expected splitting resulting from fiber ellipticity we use the known relation
\begin{equation}
  \frac{\nu_{\mathrm{split}}}{\nu_{\mathrm{FSR}}} =   \frac{\lambda}{4 \cdot \pi^2} \frac{R_1 - R_2}{R_1 R_2} 
\end{equation}
from \cite{Uphoff_2015} together with the measured radii of curvatures in the central part of the fiber mirrors from section \ref{Sec:FiberDetails} and the wavelength $\lambda = 693.7 \, \mathrm{nm}$ as used in the experiments. For fiber mirror F$_{\mathrm{B}}$ this results in $\left(\nu_{\mathrm{split}} / \nu_{\mathrm{FSR}}\right)^{\mathrm{th,F_B}} \approx 3.92\times10^{-5}$ and for fiber mirror F$_{\mathrm{A}}$ in $\left(\nu_{\mathrm{split}} / \nu_{\mathrm{FSR}}\right)^{\mathrm{th,F_A}} \approx 4.17\times10^{-5}$, respectively. Those values are in good agreement with the directly measured splitting.

\section{Birefringence-induced polarization splitting}
\label{sec:birefringence_derivation}
When the two polarization eigenmodes face birefringence in the diamond membrane, they acquire a phase difference per round trip due to the refractive index difference $\Delta n$, given by
\begin{equation}
    \Delta \varphi =2k \Delta n t_\mathrm{d}
\end{equation}
with $k=2\pi/\lambda$ the wave vector and $t_\mathrm{d}$ the diamond thickness. We measure the frequency spacing between the polarization modes in comparison to one free spectral range, defined by a round trip phase of $2\pi$. We can therefore express the splitting via
\begin{equation}
    \frac{\nu_{\mathrm{split}}}{\nu_{\mathrm{FSR}}}=\frac{\Delta \varphi}{2\pi}=\frac{2k \Delta n t_\mathrm{d}}{2\pi}= (2/\lambda) \Delta n t_\mathrm{d}.
\end{equation}
With $\delta\equiv\frac{\nu_{\mathrm{split}}}{\nu_{\mathrm{FSR}}}$, this yields the expression for $\Delta n$,
\begin{equation}
    \Delta n = \frac{\lambda \delta}{2 t_\mathrm{d}}.
\end{equation}

\section{Polarization splitting of an electronic grade sample}
\begin{figure}[tb]
    \centering
    \includegraphics[width=0.9\textwidth]{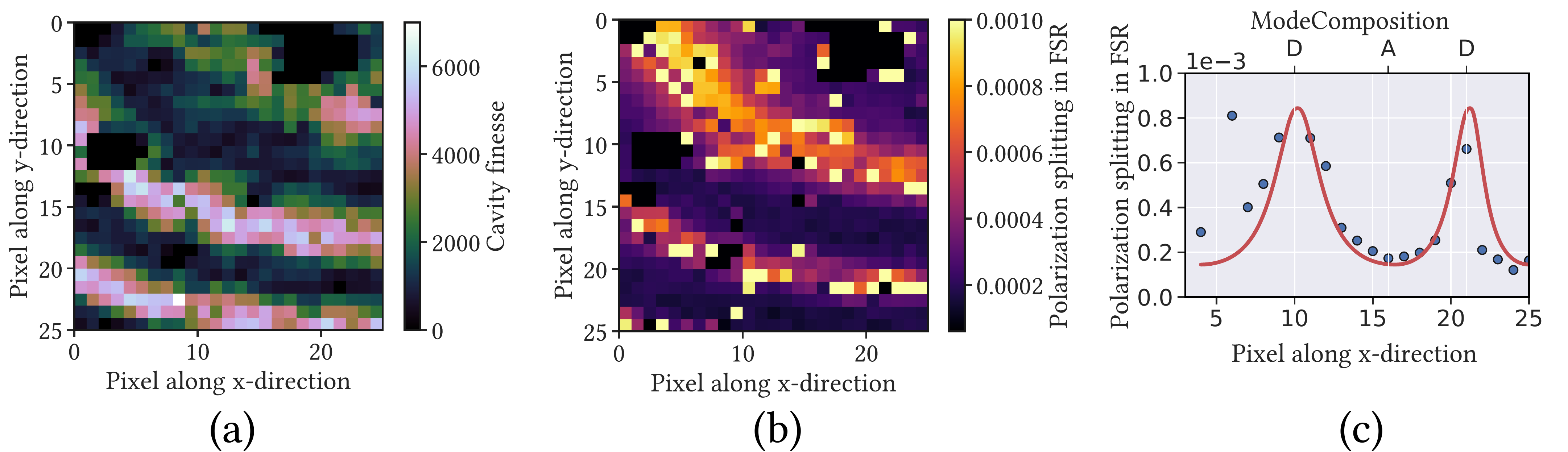}
\caption[]{Cavity finesse and polarization splitting on an electronic grade diamond membrane. \textbf{(a)} Cavity finesse scan on a sub-region of the electronic grade sample with fiber mirror F$_{\mathrm{A}}$, plane mirror M$_{\mathrm{B}}$ at a laser wavelength of $\lambda = 639.7 \, \mathrm{nm}$. \textbf{(b)} Polarization splitting of fundamental cavity modes on the same sub-region and with the same mirrors than the finesse scan in (a). \textbf{(c)} Polarization splitting for the line along $x=12$ in (b). The red line shows the same model as in the main manuscript which is consistent for different samples.}
\label{Fig6}
\end{figure}

To compare the level of birefringence of the main general grade diamond sample of our study to a sample with an expected lower level we performed a short series of measurements on a second diamond sample. This second diamond sample is an electronic grade sample that was processed as described in \cite{Heupel.2020} and bonded onto the plane mirror M$_{\mathrm{B}}$. The measurements were performed in a thinned-down region of the sample with a thickness of about $4 \, \upmu\mathrm{m}$. To save time, only a small region at a rather low resolution was investigated in lateral cavity finesse and polarization splitting measurements. As shown in figure \ref{Fig6} (a), the cavity shows finesse values between $700 - 6000$. The investigated splitting of fundamental polarization modes - depicted in figure \ref{Fig6} (b) - shows values between $2\times10^{-4}$ and $9\times10^{-4}$ and its modulation again follows strongly the cavity finesse. From the lower splitting value within regions with strong air-like character we can deduce a birefringence of $\Delta n\approx 1.6 \times10^{-5}$. Overall, we found a five-fold decrease of the maximum polarization splitting and a four-fold decrease of the birefringence on the electronic grade sample compared to the general grade sample. Figure \ref{Fig6} (c) shows that the model which was used to fit the data in the main manuscript is valid for this sample as well, showing that the polarization splitting follows the energy distribution in the cavity.

\section{Repeatability of the birefringence detection}
\begin{figure}[tb]
    \centering
    \includegraphics[width=0.9\textwidth]{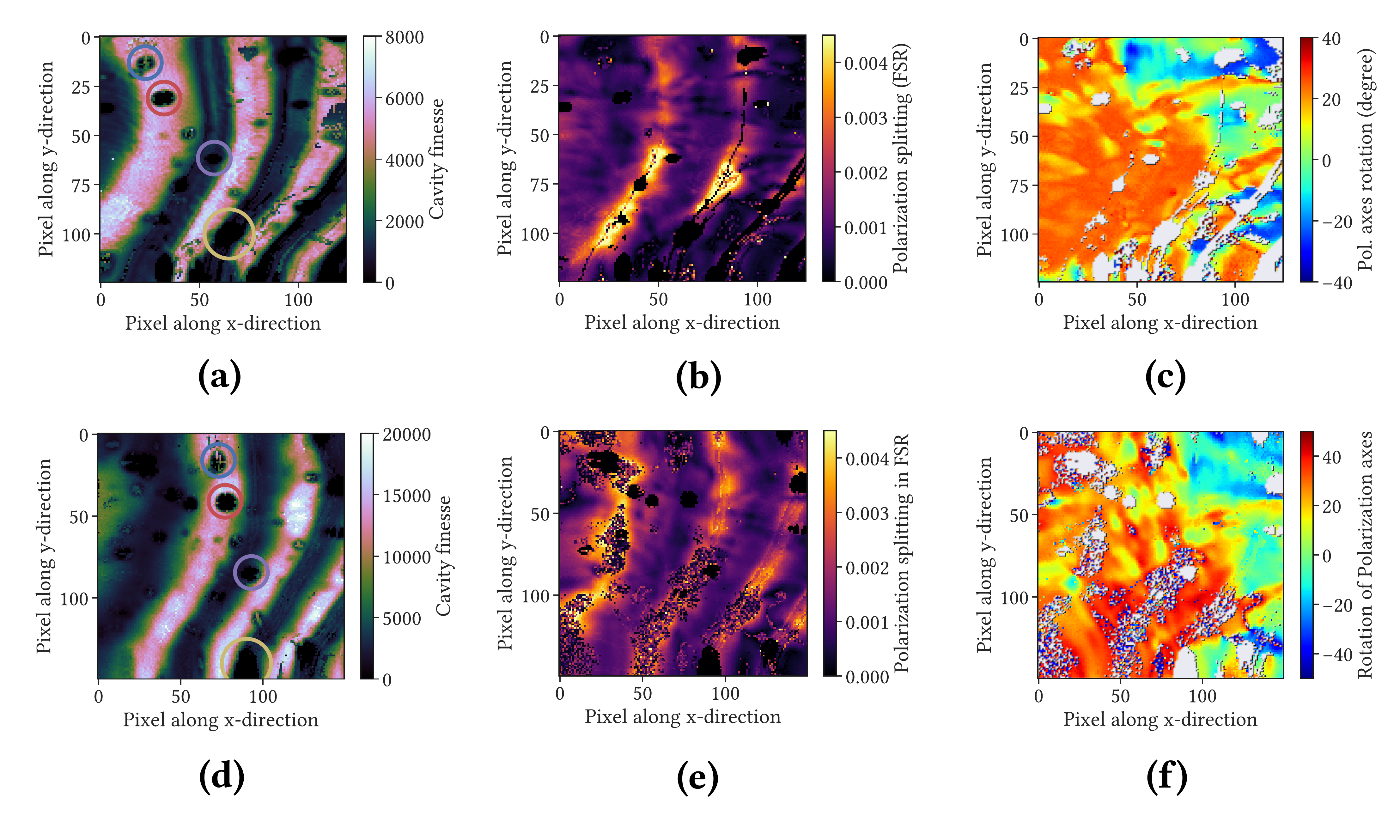}
\caption[]{Polarization splitting and rotation of the fundamental cavity modes from different cavity setups \textbf{(a)} - \textbf{(c)}: Lateral measurements of the cavity finesse and the polarization dependent splitting and rotation of fundamental cavity modes on the single-crystal sample measured with fiber mirror F$_{\mathrm{A}}$ and plane mirror M$_{\mathrm{A}}$ as already discussed in the main manuscript in figure 6. The circles in \textbf{(a)} show four distinct features in the finesse scan that can be used to determine the position mismatch compared to the second series of measurement shown in \textbf{(d) - \textbf{(f)}}. \textbf{(d) - \textbf{(f)}}:  Measurements of the cavity finesse and the polarization dependent splitting and rotation of fundamental cavity modes on almost the same position than \textbf{(a)} - \textbf{(c)} performed with a new cavity made from fiber mirror F$_{\mathrm{B}}$ and plane mirror M$_{\mathrm{A}}$.}
\label{Fig7}
\end{figure}

To exclude that the features we identified with a rotation of the cavity polarization axes induced by birefringence variations of the diamond sample can be explained by artefacts related to the cavity fiber F$_{\mathrm{A}}$ rather than the diamond sample, we decided to repeat the measurement after a rebuild of our setup with the second cavity fiber F$_{\mathrm{B}}$. In order to find the same position on the diamond sample again, we performed cavity transmission scans and used local artefacts on the membrane for orientation. As shown in figure \ref{Fig7} (d), we were able to find the position after the rebuild and moved the sample to almost the same position as the previous measurements in figure \ref{Fig7} (a). Four defects - highlighted by different colored circles in \ref{Fig7} (a) and \ref{Fig7} (d) - show that our second measurement position is slightly rotated compared to the previous one.

Figure \ref{Fig7} (e) and (f) show the repeated measurements of the polarization mode splitting and the rotation of the cavity polarization axes, respectively. Compared to the measurement with the previous setup, shown in Figure \ref{Fig7} (b) and (c), the detected features appear very similar. Especially for the measurements of the polarization axes rotation (i. e. fig \ref{Fig7} (c) and (f) ) our repeated measurements clearly indicate that indeed the features we see arise from the sample and are not related to our cavity setup in general (e.g. a bending induced polarization rotation of the cavity fiber during the scanning movement).

\section{Purcell effect and geometrical limitations}
In order to highlight the relevance of the presented cavity geometry for future applications, we gauge the suitability of such a system with a simple estimation.
For a cavity with an integrated diamond membrane of thickness $t_d$, the mode waist radius is given by \cite{SuzanneBvanDam.2018}
\begin{equation}
    w_0 = \sqrt{\frac{\lambda}{\pi}}\left(\left(t_a+\frac{t_d}{n_d}\right)\left(ROC-\left(t_a+\frac{t_d}{n_d}\right)\right)\right)^{1/4}
\end{equation}
where $t_a$ is the air gap and $ROC$ is the radius of curvature of the fiber mirror. This yields for cavity used in the main manuscript a beam waist of $w_0=1.7$\,\textmu m. The mode volume is described as
\begin{equation}
V_m = \frac{\pi}{4} w_0^2 L_\mathrm{eff}.
\end{equation}
Note that the effective cavity length $L_\mathrm{eff}$ appears in the quality factor as well and therefore cancels out when calculating the Purcell factor. A more detailed description of this can be found in ref.\cite{SuzanneBvanDam.2018}.
\begin{figure}[tb]
    \centering
    \includegraphics[width=0.9\textwidth]{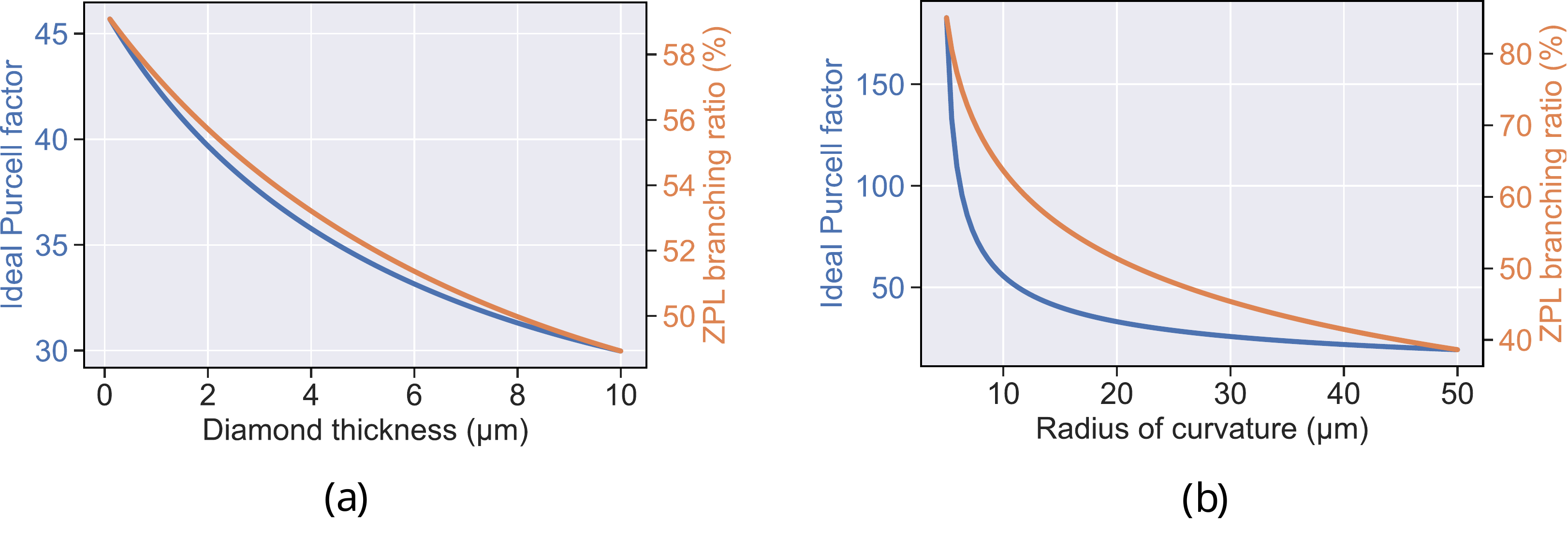}
\caption[]{Simulation of the Purcell enhancement for different cavity geometries. \textbf{(a)} Ideal Purcell factor and ZPL branching ratio for different diamond membrane thicknesses. For the simulation, coatings F$_\mathrm{A}$ and M$_\mathrm{A}$ were used with a radius of curvature of 20\,\textmu m for the fiber mirror. \textbf{(b)} Similar simulation to (a) for a fixed membrane thickness of 6\,\textmu m (as is the case in the main manuscript) and different radii of curvature.}
\label{Fig8}
\end{figure}
The ideal Purcell factor for 100-oriented diamond with NV centers located in the field antinode is then given by
\begin{equation}
    C_\mathrm{ideal} = \frac{1}{3}\frac{3}{4\pi^2}\left(\frac{\lambda}{n_d}\right)^3 \frac{Q}{V_m}
\end{equation}
where the factor $1/3$ is due to the non-ideal orientation of the NV dipole with respect to the cavity light field. Figure \ref{Fig8} shows expected Purcell factors and the respective increased branching ratio into the zero phonon line for different cavity geometries. For a cavity design consisting of the coatings of fiber mirror F$_\mathrm{A}$ and plane mirror M$_\mathrm{A}$, an increased ZPL branching ratio above 50\,\% can be reached using a fiber profile with a radius of curvature of 20\,\textmu m. Furthermore, the cavity mode volume can be minimized by decreasing the radius of curvature of the fiber mirror, as shown for a membrane thickness of 6\,\textmu m in fig.\ref{Fig8} (b). As highlighted in the main manuscript, a smaller ROC leads to increased losses. Thus, the choice of the ROC is a trade-off that becomes relevant for any membrane thickness. These estimations show that the geometries used in this work are realistic starting points for the integration of diamond membranes into fiber-based microcavities.\\
Depending on whether the light is coupled in or out of the cavity through the plane mirror or the fiber mirror, the mismatch of the fiber and the cavity modes becomes important. It is quantified by the mode matching factor $\epsilon$, given by \cite{Benedikter2017}
\begin{equation}
    \epsilon=\frac{4}{\left(\frac{w_f}{w_0}+\frac{w_0}{w_f}\right)^2+\left(\frac{s \lambda}{\pi w_0 w_f}\right)^2},
\end{equation}
where $w_f$ describes the mode field radius of the fiber mode and $s=d+d_\mathrm{mirror}$ is the optical distance between the two waists. The mode matching thereby depends mainly on the cavity length, i.e. the cavity air gap $t_a$. For short cavity lengths, the mode matching amounts to $\epsilon\approx 60$\,\% but can be improved by implementing mode matching optics to the fiber mirror, as proposed in \cite{gulati_fiber_2017}.